\documentclass[compsoc, conference, a4paper, 10pt, times]{IEEEtran}

\usepackage{booktabs}
\usepackage[colorlinks=true,urlcolor=blue]{hyperref}
\usepackage{cite}
\usepackage{amsmath,amssymb,amsfonts}
\usepackage{algorithmic}
\usepackage{graphicx}
\usepackage{textcomp}
\usepackage{bmpsize}
\usepackage[table]{xcolor}
\usepackage{lipsum}

\usepackage{cleveref}
\crefformat{section}{\S#2#1#3}
\crefformat{subsection}{\S#2#1#3}
\crefformat{subsubsection}{\S#2#1#3}
\crefrangeformat{section}{\S\S#3#1#4 to~#5#2#6}
\crefmultiformat{section}{\S\S#2#1#3}{ and~#2#1#3}{, #2#1#3}{ and~#2#1#3}
\def\BibTeX{{\rm B\kern-.05em{\sc i\kern-.025em b}\kern-.08em
    T\kern-.1667em\lower.7ex\hbox{E}\kern-.125emX}}

\usepackage{tikz}
\usetikzlibrary{trees,positioning,shapes,shadows,arrows}
\newcommand{\nop}[1]{}

\newcommand{\etal}{\emph{et~al. }}
\newcommand{\eg}{\emph{e.g., }}
\newcommand{\ie}{\emph{i.e., }}
\usepackage{array}
\usepackage{soul,xcolor}
\usepackage{multirow}
\usepackage{tabularx}
\newcolumntype{P}[1]{>{\centering\arraybackslash}p{#1}}

\definecolor{lightgray}{gray}{0.95}

\usepackage{booktabs}

\usepackage{adjustbox}
\newcolumntype{R}[2]{%
    >{\adjustbox{angle=#1,lap=\width-(#2)}\bgroup}%
    l%
    <{\egroup}%
}
\newcommand*\rot{\multicolumn{1}{R{45}{0.1em}}}

\usepackage[strict]{changepage}
\usepackage{framed}
\definecolor{formalshade}{rgb}{0.95,0.95,1}
\definecolor{darkblue}{rgb}{0.14,0.22,0.62}
\newenvironment{formal}{%
  \MakeFramed{\advance\hsize-\width\FrameRestore}%
  \noindent\hspace{-4.55pt}
  \begin{adjustwidth}{}{7pt}%
}
{%
  \end{adjustwidth}\endMakeFramed%
}

\begin{document}

\title{SoK: Explainable Machine Learning for Computer Security Applications}

\author{\IEEEauthorblockN{
Azqa Nadeem\IEEEauthorrefmark{1}, Daniël Vos\IEEEauthorrefmark{1}, Clinton Cao\IEEEauthorrefmark{1}, Luca Pajola\IEEEauthorrefmark{2}, Simon Dieck\IEEEauthorrefmark{1}, Robert Baumgartner\IEEEauthorrefmark{1}, Sicco Verwer\IEEEauthorrefmark{1}}
\IEEEauthorblockA{\IEEEauthorrefmark{1}Delft University of Technology}
\IEEEauthorblockA{\IEEEauthorrefmark{2}University of Padua}
\IEEEauthorblockA{\{azqa.nadeem, d.a.vos, c.s.cao, s.dieck, r.baumgartner-1, s.e.verwer\}@tudelft.nl, \{pajola\}@math.unipd.it}
}

\maketitle

\begin{abstract}
Explainable Artificial Intelligence (XAI) aims to improve the transparency of machine learning (ML) pipelines.
We systematize the increasingly growing (but fragmented) microcosm of studies that develop and utilize XAI methods for defensive and offensive cybersecurity tasks. 
We identify 3 cybersecurity stakeholders, \ie model users, designers, and adversaries, who utilize XAI for 4 distinct objectives within an ML pipeline, namely 1) XAI-enabled user assistance, 2) XAI-enabled model verification, 3) explanation verification \& robustness, and 4) offensive use of explanations. 
Our analysis of the literature indicates that many of the XAI applications are designed with little understanding of how they might be integrated into analyst workflows -- user studies for explanation evaluation are conducted in only 14\% of the cases.
The security literature sometimes also fails to disentangle the role of the various stakeholders, \eg by providing explanations to model users and designers while also exposing them to adversaries. Additionally, the role of model designers is particularly minimized in the security literature.
To this end, we present an illustrative tutorial for model designers, demonstrating how XAI can help with model verification. We also discuss scenarios where interpretability by design may be a better alternative.
The systematization and the tutorial enable us to challenge several assumptions, and present open problems that can help shape the future of XAI research within cybersecurity. 
\end{abstract}


\maketitle

\section{Introduction}
Security practitioners are interested in high-performing machine learning (ML) systems that can also explain their decisions \cite{rieck2008learning}. 
However, despite the unprecedented performance achieved by prevailing ML systems, they have been slow to materialize in the security industry \cite{xai_deloitte_2022,apruzzese2022role,hale2021zero}. 
This is because these systems are considered `black boxes' due to their lack of transparency --- they are notoriously difficult to understand for humans because of their complex configurations and large model sizes. In addition to the lack of understandability, their correctness and robustness can also not be easily verified. For instance, the model might learn incorrect associations (\ie spurious correlations) from the input data, giving it the illusion of being performant without being able to generalize in practice\footnote{This is often referred to as the Clever Hans phenomenon \cite{samhita2013clever}.}.
The model might also have fatal weaknesses that can be exploited by an adversary to evade detection\footnote{Adversarial machine learning studies these cases, see \eg \cite{rosenberg2021adversarial,biggio2018wild,papernot2018sok}.}. 
For the 
safety-critical environment 
of cybersecurity, the usage of such models is not ideal. In fact, black-box models are not even allowed in regulated fields unless they are supplemented with explanations \cite{6_giudici2021explainable}, \eg courts do not consider model outputs as admissible evidence unless a forensic analyst is able to justify how the output links to the case \cite{conti-for}. Moreover, the ``right to explanation'' in the GDPR AI act also makes it tricky to deploy black-box models \cite{goodman2017european}.

Explainable artificial intelligence (XAI)\footnote{The terms `explainable artificial intelligence', `explainable ML', and `interpretable ML' are used interchangeably in the literature.} has been proposed to open the proverbial `black box' by making the model internals more human understandable \cite{miller2019explanation}. 
The first mention of XAI can be traced back to van Lent \etal \cite{van2004explainable} in 2004, while the field really started growing drastically after DARPA announced its XAI program in 2014 \cite{dw2019darpa}.

In this paper, we systematize the increasingly growing microcosm of studies that develop and utilize XAI methods for security-specific target domains. We argue that the applications of XAI within cybersecurity are intrinsically different from other domains because cybersecurity works with practical use cases for safety-critical and high-stakes decision-making under adversarial settings.
The lack of explainability is also an obstacle for deployment in cybersecurity \cite{xai_deloitte_2022}. 
In fact, explainability has arguably always been a core tenet in the design of ML pipelines for security \cite{rieck2008learning,10_arp2014drebin}.
Even the seminal work by van Lent \etal uses XAI to explain the behaviour of AI-controlled entities in \textit{military simulation games} \cite{van2004explainable}.

In recent years, the security community has actively adopted XAI as a means to increase practitioners' trust \cite{samtani2020trailblazing}. 
Numerous studies have applied existing XAI methods to security applications \cite{12_alperin2020improving,13_antwarg2021explaining}.  However, recent studies have recognized their shortcomings in 
addressing the unique pain points of the security domain \cite{paredes2021importance}. 
To this end, several security-specific XAI methods \cite{15_guo2018lemna,16_han2021deepaid,17_yang2021cadeusenix}, and evaluation criteria \cite{warnecke2020evaluating,ganz2021explaining} have been proposed.

\begin{figure*}[t]
    \centering
    \includegraphics[width = 0.9\linewidth]{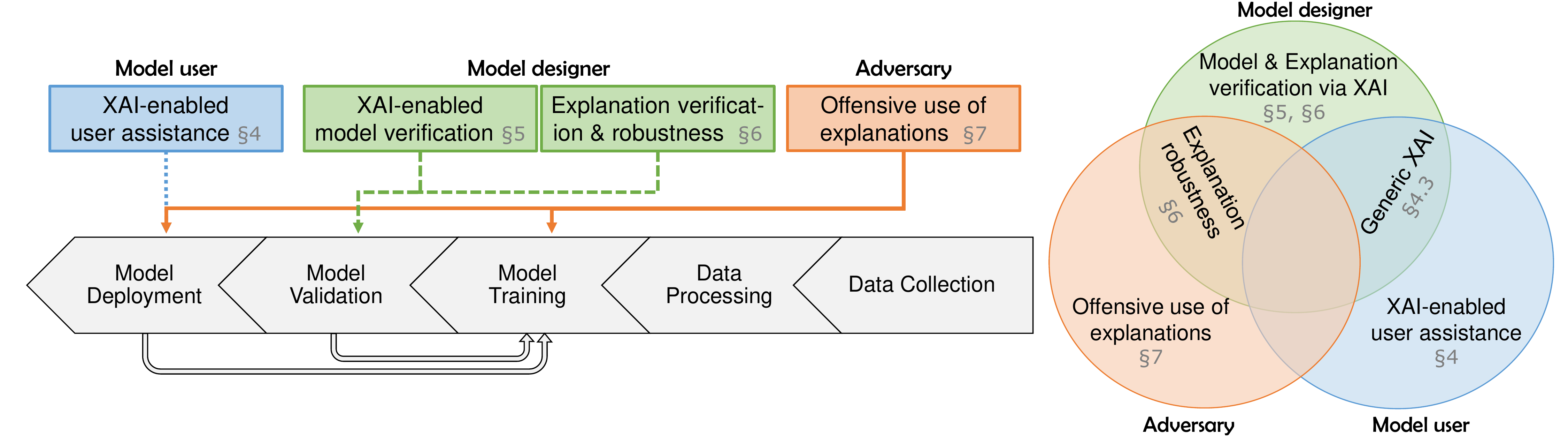}
    \caption{The interplay between application objectives and an ML pipeline (Left); and stakeholders and application objectives (Right).}
    \label{fig:cyber-sec}
\end{figure*}

These recent developments have made XAI research within cybersecurity a fast-growing field: while there were only 42 articles about \textit{`explainability'}, \textit{`learning'}, and \textit{`cybersecurity'} in 2015, that number has since skyrocketed to 2600+ in 2021, according to Google Scholar. 
This literature is fragmented across several research communities (including ML, security, graphics, and software engineering) with no unified overview. Additionally, existing works often use different terminologies interchangeably, \eg explicable, accountable, transparent, and understandable, making it more difficult to find relevant literature.

To the best of our knowledge, this is the first SoK on explainable ML for cybersecurity\footnote{Although Hariharan \etal \cite{hariharan2021explainable} present a short survey on XAI for cybersecurity, it covers only a small fraction of the literature. Moreover, the Explainable Security (XSec) framework proposed by Vigano \etal \cite{vigano2020explainable} is a non-conventional take on explainability, and does not embed the traditional XAI concepts within the security context.
}. 
By taking a step back, we synthesize insights from the vast body of fragmented literature and identify open areas to stimulate further research in this field.

Following the XAI definitions set forth by Roscher \etal \cite{roscher2020explainable}, we identify three cybersecurity stakeholders, \ie \textit{model users}, \textit{model designers}, and \textit{adversaries} who utilize XAI for four distinct objectives within the security literature: (i) XAI-enabled user assistance, (ii) XAI-enabled model verification, (iii) explanation verification \& robustness, and (iv) offensive use of explanations.
The interplay between the stakeholders, objectives, and the stages of a typical ML pipeline are given in Figure \ref{fig:cyber-sec}. 
Particularly, the stakeholders remain central throughout our discussions. We further categorize the literature \textit{w.r.t} the targeted security domain (\eg intrusion detection), ML model, and XAI method.
This taxonomy serves as a guide for finding related literature on XAI for cybersecurity.

After carefully reviewing 300+ papers, we found that XAI has been most commonly used for providing decision support to model users -- 58\% of the works are classified under \textit{XAI-enabled user assistance}. User evaluation is a critical aspect of these studies to ensure that they are usable, and are sufficiently aligned with existing analyst workflows. However, user studies are conducted in only 14\% of the cases, which is alarming since these methods aim to work directly with model users. We propose ideas for mitigating the lack of user studies in \cref{sec:discussion}.

In addition, the stakeholders we identify have different competencies, and thus require tailored explanations \cite{blumreiter2019towards}, \eg model designers are typically experts in ML while model users are not. However, we identify several cases that either do not distinguish between model users and designers or do not specify any stakeholder. In contrast, model users and adversaries interact with the explanations in similar ways, but with opposing intent, requiring special manoeuvres to limit adversary access.

Furthermore, the role of model designers is substantial in cybersecurity for ensuring the security of the model and its explanations. Yet, the reviewed literature only provides decision support to model designers in 22.3\% of the cases. We tease out the role of model designers in \cref{sec:usecase}: we present a walk-through tutorial of how model designers can utilize XAI to detect and remove spurious features in a network attack detection scenario. 
The tutorial serves as a practical and easy starting point for security practitioners by showing three concrete ways to debug a black-box model via XAI. We show cases where the explanations are helpful, and cases where model designers may draw misleading conclusions instead. We discuss what can go wrong when explaining black-box models, and advocate for interpretability by design.

\noindent\textbf{Organization:} In \cref{sec:method} and \cref{sec:tax}, we describe the scope and the proposed taxonomy. 
In \cref{sec:decision-support}-\cref{sec:offensive-xai}, we elaborate on the main takeaways from the reviewed literature. In \cref{sec:usecase}, we demonstrate how model designers can use XAI to debug their models.
In \cref{sec:discussion}, we present open problems and recommendations for further XAI research within cybersecurity. 

\section{Background and Methodology}\label{sec:method}

\begin{figure}[t]
    \centering
    \includegraphics[width=\linewidth]{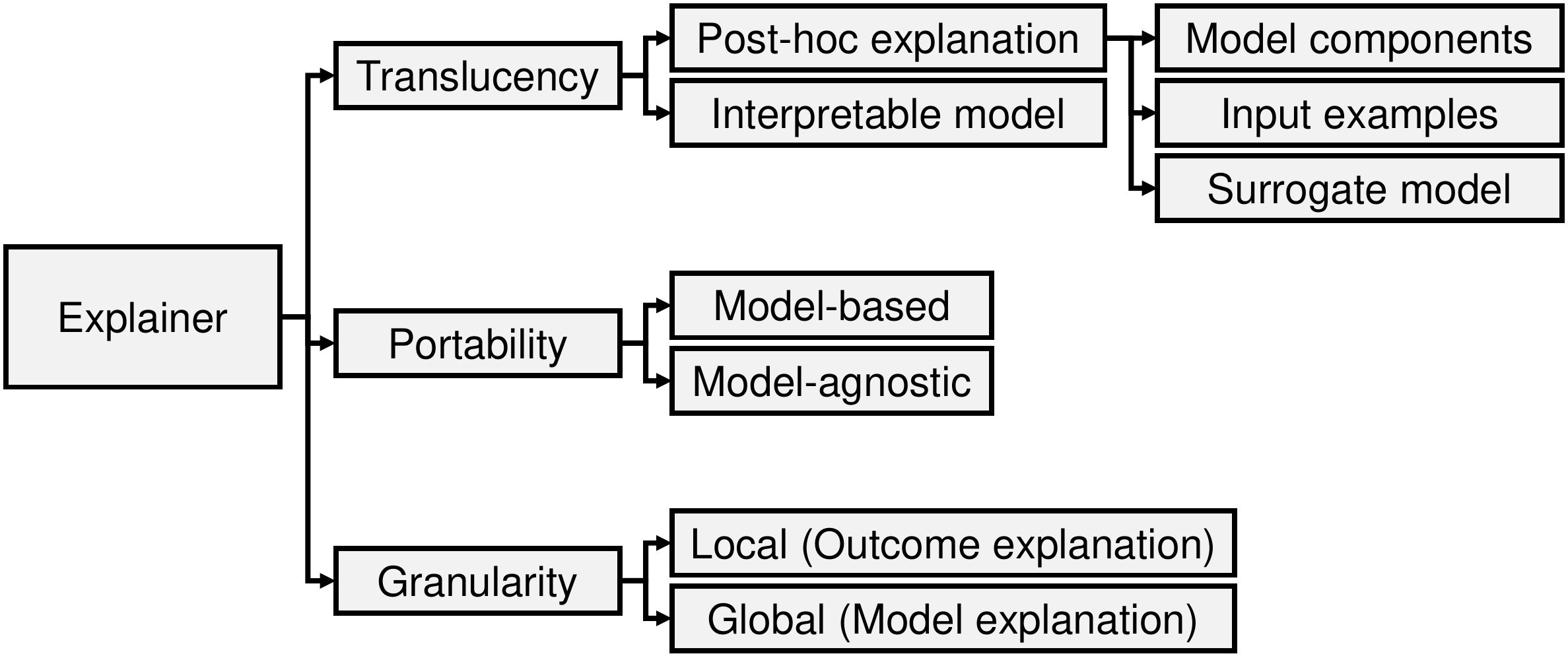}
    \caption{XAI ontology showing the key concepts within XAI.}
    \label{fig:XAI-ontology}
\end{figure}

\noindent{\textbf{Explainable machine learning.}} ML pipelines either use white-box models, which are inherently \textit{interpretable}, or use black-box models that are explained via \textit{post-hoc explainability}. For instance, linear regression and decision trees are considered white-box, while neural networks and random forests are considered black-box. 
The output of a post-hoc explainability method is either an \textit{interpretable surrogate model} that approximates the black-box model, or is an explanation of the black-box model in terms of \textit{model components} (\eg feature importance) or \textit{input examples} (\eg counterfactuals).
Additionally, an explanation can either elucidate how the model prediction is affected by a single data point (\textit{local methods}), or by all data points (\textit{global methods}). Note that interpretable models provide both local and global explanations.
Furthermore, most post-hoc methods can be applied to any ML model, making them \textit{model-agnostic}, while interpretable models are also referred to as \textit{model-based explanations}. Figure \ref{fig:XAI-ontology} shows the various XAI terminologies, adapted from \cite{adadi2018peeking}.

\noindent{\textbf{Method and Scoping.}}
We synthesize available literature that uses XAI for (offensive and defensive) cybersecurity tasks. To this aim, we collect relevant literature, apply a reflexive thematic analysis \cite{braun2006using} to construct a taxonomy based on common themes (\ie application objectives), and classify the literature into those themes. The literature was collected by seven researchers. Each paper was investigated by at least two researchers independently and discussed with all authors during weekly meetings. The code books were updated as new categories emerged.

We collected peer-reviewed literature that has used explainable models to address cybersecurity problems since 2014, \ie post-DARPA, by searching popular scientific repositories (\eg IEEE Xplore) and top security conference proceedings (\eg Usenix). We used known search terms, 
\eg `\textit{explainable}', `\textit{interpretable}', `\textit{artificial intelligence}', `\textit{cybersecurity}', `\textit{robust}', `\textit{offensive}', `\textit{attacks}', and `\textit{trustworthiness}'.
To handle the fragmented literature, we expanded our search to include synonyms of the search terms at smaller security and non-security venues (see appendix \ref{selected-lit} for the full list of venues). We also included older popular works that try to explain their model without explicitly using XAI, see \eg \cite{10_arp2014drebin,cui2007discoverer,26_cho2010inference}. After carefully reviewing 300+ papers, we select 75 cybersecurity studies to build the taxonomy. Since it is impossible to cover all the available literature in the limited space, we chose representative works from each problem area. As such, there is some overlap with usable security, safety, and robustness literature, but we mainly focus on the use of XAI within cybersecurity. 

\section{Systematization}\label{sec:tax}

Given an ML pipeline from data collection to model deployment, explainable ML is applied once a model becomes available. In the literature, XAI has been used to explain the model output to a human, either for supporting them in decision-making or to understand whether the model works as intended. In addition, the adversarial threat landscape of cybersecurity suggests that XAI can also be used by an adversary to gain actionable information about the model in order to strengthen their attacks. This implies the existence of three stakeholders who interact with different phases of an ML pipeline and accomplish distinct objectives using XAI. We classify the literature based on the stakeholders, application objectives, target domain, model and explainer class. Figure \ref{fig:cyber-sec} shows an overview of the stakeholder objectives that can be accomplished by applying XAI on a typical ML pipeline.

\noindent{\textbf{Stakeholders.}} We identify three stakeholders (explainees) who have different intents and expertise, and thus consume explanations for distinct objectives, even when interacting with the same ML model. The explanations are hence tailored to the specific needs of the stakeholders.

\textit{\textbf{a) Model user}} is defined as a broad class of personnel who utilize the ML pipeline to improve the defence capacity of an organization, such as an analyst, developer, operator, domain expert, practitioner, or end-user. To this aim, a model user utilizes XAI techniques to better understand the output of a deployed model and make informed decisions, \eg a malware analyst uses explanations to gain insights into why a binary was classified as malicious \cite{74_mathews2019explainable}.

\textit{\textbf{b) Model designer}} is responsible for engineering the ML pipeline used for a security application, and consequently has a more intimate relationship with the model. A model designer utilizes XAI techniques during model training and validation to verify that the model works as intended, \eg a malware analyst uses explanations to investigate the causes of misclassifications, and to ensure that the model employs meaningful features \cite{63_becker2020interpretable}. Moreover, since the ML pipeline exists in an adversarial threat landscape, the model designer also ensures the safety and robustness of the model and its explanations \cite{29_alvarez2018robustness}.

\textit{\textbf{c) Adversary}} is a human or an automated agent (malware) that intends to harm an organization by compromising the ML pipeline. An adversary exploits XAI techniques to formulate more efficient attacks, \eg by discovering weaknesses in the model \cite{31_kuppa2021adversarial}. In addition, an adversary may attack the XAI component of the ML pipeline itself to alter the generated explanations \cite{30_ghorbani2019interpretation}. Depending on the attacker model, the adversary may interact with the explanations either during model training or after the model is deployed.

\noindent{\textbf{Application objectives.}} We classify the literature under four application objectives based on the intended use of the XAI technique --- XAI is used to provide decision support to model users in (1); model designers in (2) \& (3); and adversaries in (4).

\textit{\textbf{(1) XAI-enabled user assistance}} covers techniques that are developed and utilized to support \textit{model users} in making informed decisions, usually with the help of visual analytics dashboards. 
The explanations are meant to give control back to the user by helping them understand the model \cite{33_nadeem2021beyond}, and providing additional insights regarding the input data \cite{34_Cao2022}. Since it is the model designers who typically develop the explanations for model users, it is essential to include model users during the evaluation process to understand the explanation efficacy.   

\textit{\textbf{(2) XAI-enabled model verification}} studies techniques that are developed and utilized to help \textit{model designers} debug and validate the correctness of the ML model. These explanations are usually more technical in nature. In the literature, XAI has primarily been used to discover spurious/faulty features by investigating a given black-box model using, \eg feature importance \cite{63_becker2020interpretable,84_chakraborty2021deep} or surrogate models \cite{16_han2021deepaid,42_dolejvs2022interpretability}.

\textit{\textbf{(3) Explanation verification \& robustness}} studies techniques that are developed and utilized to help \textit{model designers} debug and validate the correctness \& robustness of the XAI component in the ML pipeline. These methods focus on testing the correctness of post-hoc explanations under natural settings \cite{87_lin2020you}, and the robustness of explanations produced by post-hoc methods \cite{29_alvarez2018robustness} and interpretable models \cite{44_vos2021efficient} under adversarial settings \cite{86_dombrowski2019explanations}.

\textit{\textbf{(4) Offensive use of explanations}} studies how \textit{adversaries} can exploit insights provided by XAI techniques for enhancing their capabilities, \eg i) by using explanations to compromise the privacy of the model, and ii) by using explanations to compromise the integrity and availability of the model. These attacks can be deployed in the model training phase (\eg poisoning attacks \cite{45_severi2021explanation}), and model deployment phase (\eg privacy attacks \cite{46_zhao2021exploiting}).

\noindent{\textbf{Target domain.}} We further classify the literature according to the cybersecurity target domain. In terms of defensive security domains, we cover: \textit{malware detection}, \textit{anomaly detection}, \textit{intrusion detection}, \textit{alert management}, \textit{vulnerability discovery}, \textit{asset prioritization}, \textit{phishing detection}, \textit{reverse engineering}, \textit{traffic classification}, and \textit{privacy protection}. In terms of offensive security domains, we cover: \textit{privacy attacks} (\eg \textit{membership inference}, \textit{model inversion}, and \textit{model extraction}), \textit{poisoning attacks} (\eg \textit{backdoor injection}), and \textit{evasion attacks} (\eg \textit{test-time adversarial perturbations}). 
The papers that address generic non-security concepts, such as \textit{anomaly detection} are further categorized according to the data sources they use, \eg image, binary, and network traffic. To the best of our knowledge, this is the first SoK to cover such a broad range of target domains. 

\noindent{\textbf{Model \& explainer class.}} Finally, we specify the ML models and XAI techniques used in the literature. The models are grouped according to the algorithm and the input data type accepted by the model (\eg tabular, images). The models are further classified coarsely as either black-box or white-box models, following the consensus of the ML community, see Table \ref{tab:model-classes} for the model code book.
The XAI techniques (called explainers henceforth) are categorized according to their underlying mechanism (\eg model components, examples, surrogate), see Table \ref{tab:explainer-classes} for the explainer code book.

Table \ref{tab:full-taxonomy} provides a summary of the reviewed literature, which also showcases the co-occurrence of certain models and explainers. Note that the classification in Table \ref{tab:model-classes} reflects the general level of understanding provided by the model class, while Table \ref{tab:full-taxonomy} shows the actual usage of the model: some studies explicitly treat white-box models as black-box for a model-free approach, see \eg \cite{12_alperin2020improving,41_de2021trust}. Other studies utilize an incomprehensible feature set (\eg by replacing feature names with integers), turning an interpretable model into a black box, see \eg \cite{47_karn2020cryptomining}. 

The overview also helps us identify the misleading usage of certain terminologies. For instance, some works report their methods as being `interpretable' while utilizing post-hoc explainers for black-box models, see \eg \cite{16_han2021deepaid,48_parrainterpretable2022ndss,49_zhang2020interpretable}. Strictly speaking, black-box models cannot be interpretable \cite{roscher2020explainable}. Thus, we categorize such works under post-hoc explainability. Note that it is possible to have an interpretable model that also uses a post-hoc explainer, but when a black-box model is explained via an interpretable model, it is called a surrogate model. 

\begin{table}[t]
\rowcolors{2}{lightgray}{}
\caption{Code book for ML model classes. `w' represents white-box, and `b' represents black-box models.}
\label{tab:model-classes}
\centering
\resizebox{\columnwidth}{!}{%
\setlength{\tabcolsep}{2pt}
\tiny
\begin{tabular}{llll}
\toprule
\textbf{\textbf{Model class}} & \textbf{\textbf{Machine learning algorithms}} & \textbf{w} & \textbf{b} \\ \midrule
CNN & \begin{tabular}[c]{@{}l@{}}Convolutional neural networks for image data, \eg ResNet, \\VGGnet, RPN, and inception network\end{tabular} &  & \ \textbullet \ \\
DNN & \begin{tabular}[c]{@{}l@{}}Deep neural networks for tabular data, \eg MLP, and \\auto-encoder\end{tabular} &  & \ \textbullet \ \\
GNN & \begin{tabular}[c]{@{}l@{}}Graph neural networks, \eg GCN, and graph attention network\end{tabular} &  & \ \textbullet \ \\
SeqNN & \begin{tabular}[c]{@{}l@{}}Sequential neural networks, \eg RNN, LSTM, and transformers\end{tabular} &  & \ \textbullet \ \\
Kernel-SVM & \begin{tabular}[c]{@{}l@{}}Support vector machine with non-linear kernel\\\end{tabular} &  & \ \textbullet \ \\
Ensemble & \begin{tabular}[c]{@{}l@{}}Ensemble of models, \eg random forest, gradient boosting \\trees, and neural network ensembles\end{tabular} &  & \ \textbullet \ \\
LM & \begin{tabular}[c]{@{}l@{}}Linear models, \eg logistic (rule) regression, linear rank \\regression, and linear SVM\end{tabular} & \ \textbullet \ &  \\
RBC & \begin{tabular}[c]{@{}l@{}}Rule-based classifiers, \eg decision trees, regular expressions, \\and BRCG\\\end{tabular} & \ \textbullet \ &  \\
NB & \begin{tabular}[c]{@{}l@{}}Naive Bayes and its gaussian variant\\\end{tabular} & \ \textbullet \ &  \\
Automata & \begin{tabular}[c]{@{}l@{}}Abstract computing machines, \eg Markov chains, and prob-\\abilistic deterministic finite automata\end{tabular} & \ \textbullet \ &  \\
kNN & \begin{tabular}[c]{@{}l@{}}K-nearest neighbors\\\end{tabular} & \ \textbullet \ & \ \textbullet \ \\
Unsupervised & \begin{tabular}[c]{@{}l@{}}Clustering algorithms, \eg HDBSCAN, kmeans, with(out) dim-\\ensionality reduction, \eg self-organizing maps, PCA, t-SNE\end{tabular} & \ \textbullet \ & \ \textbullet \ \\ \bottomrule
\end{tabular}}
\end{table}

\begin{table}[t]
\rowcolors{2}{lightgray}{}
\caption{Code book for explainer classes.}
\label{tab:explainer-classes}
\centering
\resizebox{\columnwidth}{!}{%
\setlength{\tabcolsep}{2pt}
\tiny
\begin{tabular}{ll}
\toprule
\textbf{Explainer class} & \textbf{Explanation methods} \\ \midrule
SHAP & \begin{tabular}[c]{@{}l@{}}SHAP  and its variants, \eg kernelSHAP \\\end{tabular} \\ 
LIME & \begin{tabular}[c]{@{}l@{}}LIME and its variants, \eg graphLIME \\\end{tabular} \\ 
LEMNA & \begin{tabular}[c]{@{}l@{}}Non-linear LIME variant for security applications\\\end{tabular} \\ 
GNNExplainer & \begin{tabular}[c]{@{}l@{}}Explanation method for graph neural networks\\\end{tabular} \\ 
Grad-based & \begin{tabular}[c]{@{}l@{}}Gradient-based methods, \eg GradCAM, saliency map, \\integrated gradients, and layer-wise relevance propagation\\\end{tabular} \\ 
Activation & \begin{tabular}[c]{@{}l@{}}Neuron activations, activation maps and attention\\\end{tabular} \\ 
Importance & \begin{tabular}[c]{@{}l@{}}Feature importance computed using tree-based splitting, feature \\permutation, and SOM-based dimensionality reduction \\\end{tabular}\\ 
Exemplar & \begin{tabular}[c]{@{}l@{}}Example-based explanations, \eg kNN, prototypes, protoDash \\\end{tabular}\\ 
Contrastive & \begin{tabular}[c]{@{}l@{}}Contrastive explanations, \eg counterfactuals\\\end{tabular}\\ 
Anomaly-score & \begin{tabular}[c]{@{}l@{}}Custom metric capturing deviation from normalcy, \eg decoder\\ reconstruction loss\\\end{tabular}\\ 
Visual-explanation & \begin{tabular}[c]{@{}l@{}}Explanation based on visualizing model components or model \\output for human perception\\\end{tabular}\\ 
Sur-RBC & \begin{tabular}[c]{@{}l@{}}Surrogate rule-based classifiers, \eg decision trees, decision \\lists, and rule sets\\\end{tabular} \\
Sur-Mixture & \begin{tabular}[c]{@{}l@{}}Surrogate mixture linear regression model\\\end{tabular} \\
Sur-Automata & \begin{tabular}[c]{@{}l@{}}Surrogate automaton model\\\end{tabular} \\ \bottomrule
\end{tabular}}
\end{table}

\begin{table*}[h!]
    \centering
    \rowcolors{2}{lightgray}{}
    \caption{Summary of XAI literature within cybersecurity. Rows are ordered \textit{w.r.t} Objectives, Target domain, and Year.}
    \setlength{\tabcolsep}{2.6pt}
    \tiny
\begin{tabular}{lrl|ccc|cccc|cc|cccccccccccc|cccccccccccccc}
\toprule
\textbf{} & \textbf{} &  & \multicolumn{3}{c}{Stakeholder} & \multicolumn{4}{c}{Objectives} &  & \multicolumn{1}{l}{Type} & \multicolumn{12}{c}{Models} & \multicolumn{14}{c}{Explainers} \\
Ref. & Year & Target domain & \rot{User} & \rot{Designer} & \rot{Adversary} & \rot{User assist.} & \rot{Model verif.} & \rot{Explanation verif.} & \rot{Offensive use} & \rot{Interpretable} & \rot{Post-hoc} & \rot{CNN} & \rot{DNN} & \rot{GNN} & \rot{SeqNN} & \rot{Kernel-SVM} & \rot{Ensemble} & \rot{LM} & \rot{RBC} & \rot{NB} & \rot{Automata} & \rot{kNN} & \rot{Unsupervised} & \rot{SHAP} & \rot{LIME} & \rot{LEMNA} & \rot{GNNExplainer} & \rot{Grad-based} & \rot{Activation} & \rot{Importance} & \rot{Exemplar} & \rot{Contrastive} & \rot{Anomaly-score} & \rot{Visual-explanation} & \rot{Sur-RBC} & \rot{Sur-Mixture} & \rot{Sur-Automata}   \hspace{0.4cm} \\ \midrule
\cite{51_sopan2018building} & 2018 & \multicolumn{1}{l|}{Alert management} & \ \textbullet \ &  & \multicolumn{1}{c|}{} & \ \textbullet \ &  &  & \multicolumn{1}{c|}{} &  & \multicolumn{1}{c|}{\checkmark} &  &  &  &  &  & \checkmark &  &  &  &  &  & \multicolumn{1}{c|}{} &  &  &  &  &  &  & \checkmark &  &  &  &  & \checkmark &  &  \\
\cite{35_nadeem2021alert} & 2021 & \multicolumn{1}{l|}{Alert management} & \ \textbullet \ &  & \multicolumn{1}{c|}{} & \ \textbullet \ &  &  & \multicolumn{1}{c|}{} & \checkmark & \multicolumn{1}{c|}{} &  &  &  &  &  &  &  &  &  & \checkmark &  & \multicolumn{1}{c|}{} &  &  &  &  &  &  &  &  &  &  &  &  &  &  \\
\cite{52_robert_deepcase} & 2022 & \multicolumn{1}{l|}{Alert management} & \ \textbullet \ &  & \multicolumn{1}{c|}{} & \ \textbullet \ &  &  & \multicolumn{1}{c|}{} &  & \multicolumn{1}{c|}{\checkmark} &  &  &  & \checkmark &  &  &  &  &  &  &  & \multicolumn{1}{c|}{} &  &  &  &  &  & \checkmark &  &  &  &  &  &  &  &  \\
\cite{53_nyre2022considerations} & 2022 & \multicolumn{1}{l|}{Alert management} & \ \textbullet \ &  & \multicolumn{1}{c|}{} & \ \textbullet \ &  &  & \multicolumn{1}{c|}{} &  & \multicolumn{1}{c|}{\checkmark} &  &  &  &  &  & \checkmark &  &  &  &  &  & \multicolumn{1}{c|}{} & \checkmark &  &  &  &  &  &  &  &  &  &  &  &  &  \\
\cite{55_lin2018tabor} & 2018 & \multicolumn{1}{l|}{Anomaly detection (sensor)} & \ \textbullet \ &  & \multicolumn{1}{c|}{} & \ \textbullet \ &  &  & \multicolumn{1}{c|}{} & \checkmark & \multicolumn{1}{c|}{} &  &  &  &  &  &  &  &  &  & \checkmark &  & \multicolumn{1}{c|}{} &  &  &  &  &  &  &  &  &  &  &  &  &  &  \\
\cite{72_brown2018recurrent} & 2018 & \multicolumn{1}{l|}{Anomaly detection (syslogs)} & \ \textbullet \ &  & \multicolumn{1}{c|}{} & \ \textbullet \  &  &  & \multicolumn{1}{c|}{} &  & \multicolumn{1}{c|}{\checkmark} &  &  &  & \checkmark &  &  &  &  &  &  &  & \multicolumn{1}{c|}{} &  &  &  &  &  & \checkmark &  &  &  &  &  &  &  &  \\
\cite{47_karn2020cryptomining} & 2020 & \multicolumn{1}{l|}{Anomaly detection (syslogs)} & \ \textbullet \ &  & \multicolumn{1}{c|}{} & \ \textbullet \ &  &  & \multicolumn{1}{c|}{} &  & \multicolumn{1}{c|}{\checkmark} &  & \checkmark &  & \checkmark &  & \checkmark &  & \checkmark &  &  &  & \multicolumn{1}{c|}{} & \checkmark & \checkmark &  &  &  &  &  & \checkmark &  &  &  &  &  &  \\
\cite{54_ardito2021artificial} & 2021 & \multicolumn{1}{l|}{Anomaly detection (network)} & \ \textbullet \ &  & \multicolumn{1}{c|}{} & \ \textbullet \ &  &  & \multicolumn{1}{c|}{} &  & \multicolumn{1}{c|}{\checkmark} &  & \checkmark &  &  &  &  &  &  &  &  &  & \multicolumn{1}{c|}{} &  &  &  &  &  &  &  &  &  & \checkmark &  &  &  &  \\
\cite{71_ardito2021revisiting} & 2021 & \multicolumn{1}{l|}{Anomaly detection (sensor)} & \ \textbullet \ &  & \multicolumn{1}{c|}{} & \ \textbullet \  &  &  & \multicolumn{1}{c|}{} & \checkmark & \multicolumn{1}{c|}{\checkmark} &  &  &  &  & \checkmark & \checkmark &  & \checkmark &  &  & \checkmark & \multicolumn{1}{c|}{} &  &  &  &  &  &  & \checkmark &  &  &  &  &  &  &  \\
\cite{56_Hwang2021esfd} & 2021 & \multicolumn{1}{l|}{Anomaly detection (sensor)} & \ \textbullet \ &  & \multicolumn{1}{c|}{} & \ \textbullet \ &  &  & \multicolumn{1}{c|}{} &  & \multicolumn{1}{c|}{\checkmark} &  &  &  &  &  & \checkmark &  &  &  &  &  & \multicolumn{1}{c|}{} & \checkmark &  &  &  &  &  &  &  &  &  &  &  &  &  \\
\cite{16_han2021deepaid} & 2021 & \multicolumn{1}{l|}{Anomaly detection (network)} & \ \textbullet \ & \ \textbullet \ & \multicolumn{1}{c|}{} & \ \textbullet \ & \ \textbullet \ &  & \multicolumn{1}{c|}{} &  & \multicolumn{1}{c|}{\checkmark} & \checkmark &  & \checkmark & \checkmark &  &  &  &  &  &  &  & \multicolumn{1}{c|}{} &  &  &  &  &  &  &  &  &  &  &  &  &  & \checkmark \\
\cite{34_Cao2022} & 2022 & \multicolumn{1}{l|}{Anomaly detection (network)} & \ \textbullet \ &  & \multicolumn{1}{c|}{} & \ \textbullet \ &  &  & \multicolumn{1}{c|}{} & \checkmark & \multicolumn{1}{c|}{} &  &  &  &  &  &  &  &  &  & \checkmark &  & \multicolumn{1}{c|}{} &  &  &  &  &  &  &  &  &  &  &  &  &  &  \\
\cite{39_li2021explainable} & 2021 & \multicolumn{1}{l|}{Asset prioritization} & \ \textbullet \ &  & \multicolumn{1}{c|}{} & \ \textbullet \  &  &  & \multicolumn{1}{c|}{} &  & \multicolumn{1}{c|}{\checkmark} & \checkmark &  &  & \checkmark &  &  &  &  &  &  &  & \multicolumn{1}{c|}{} &  & \checkmark &  &  &  &  &  &  &  &  &  &  &  &  \\
\cite{6_giudici2021explainable} & 2021 & \multicolumn{1}{l|}{Asset prioritization} & \ \textbullet \ &  & \multicolumn{1}{c|}{} & \ \textbullet \ &  &  & \multicolumn{1}{c|}{} & \checkmark & \multicolumn{1}{c|}{} &  &  &  &  &  &  & \checkmark &  &  &  &  & \multicolumn{1}{c|}{} &  &  &  &  &  &  &  &  &  &  &  &  &  &  \\
\cite{100_holder2021explainable} & 2021 & \multicolumn{1}{l|}{Asset prioritization} & \ \textbullet \ &  & \multicolumn{1}{c|}{} & \ \textbullet \ &  &  & \multicolumn{1}{c|}{} & \checkmark & \multicolumn{1}{c|}{\checkmark} &  &  &  &  &  &  &  & \checkmark &  &  &  & \multicolumn{1}{c|}{\checkmark} &  &  &  &  &  &  &  &  &  &  & \checkmark &  &  &  \\
\cite{57_szczepanski2020achieving} & 2020 & \multicolumn{1}{l|}{Intrusion detection} & \ \textbullet \ &  & \multicolumn{1}{c|}{} & \ \textbullet \ &  &  & \multicolumn{1}{c|}{} &  & \multicolumn{1}{c|}{\checkmark} &  & \checkmark &  &  &  &  &  &  &  &  &  & \multicolumn{1}{c|}{} &  &  &  &  &  &  &  &  &  &  &  & \checkmark &  &  \\
\cite{58_robert_frameworkwithshap} & 2020 & \multicolumn{1}{l|}{Intrusion detection} & \ \textbullet \ & \ \textbullet \ & \multicolumn{1}{c|}{} & \ \textbullet \ & \ \textbullet \ &  & \multicolumn{1}{c|}{} &  & \multicolumn{1}{c|}{\checkmark} &  & \checkmark &  &  &  &  &  &  &  &  &  & \multicolumn{1}{c|}{} & \checkmark &  &  &  &  &  &  &  &  &  &  &  &  &  \\
\cite{13_antwarg2021explaining} & 2021 & \multicolumn{1}{l|}{Intrusion detection} & \ \textbullet \ &  & \multicolumn{1}{c|}{} & \ \textbullet \ &  &  & \multicolumn{1}{c|}{} &  & \multicolumn{1}{c|}{\checkmark} &  & \checkmark &  &  &  &  &  &  &  &  &  & \multicolumn{1}{c|}{} & \checkmark &  &  &  &  &  &  &  &  &  &  &  &  &  \\
\cite{59_mahbooba2021explainable} & 2021 & \multicolumn{1}{l|}{Intrusion detection} & \ \textbullet \ &  & \multicolumn{1}{c|}{} & \ \textbullet \ &  &  & \multicolumn{1}{c|}{} & \checkmark & \multicolumn{1}{c|}{} &  &  &  &  &  &  &  & \checkmark &  &  &  & \multicolumn{1}{c|}{} &  &  &  &  &  &  &  &  &  &  &  &  &  &  \\
\cite{60_liu2021faixid} & 2021 & \multicolumn{1}{l|}{Intrusion detection} & \ \textbullet \ &  & \multicolumn{1}{c|}{} & \ \textbullet \ &  &  & \multicolumn{1}{c|}{} & \checkmark & \multicolumn{1}{c|}{\checkmark} &  &  &  &  &  &  & \checkmark & \checkmark &  &  &  & \multicolumn{1}{c|}{} &  &  &  &  &  &  &  & \checkmark & \checkmark &  &  &  &  &  \\
\cite{48_parrainterpretable2022ndss} & 2022 & \multicolumn{1}{l|}{Intrusion detection} & \ \textbullet \ &  & \multicolumn{1}{c|}{} & \ \textbullet \ &  &  & \multicolumn{1}{c|}{} &  & \multicolumn{1}{c|}{\checkmark} &  &  &  & \checkmark &  &  &  &  &  &  &  & \multicolumn{1}{c|}{} &  &  &  &  &  & \checkmark &  &  &  &  &  &  &  &  \\
\cite{33_nadeem2021beyond} & 2021 & \multicolumn{1}{l|}{Malware analysis} & \ \textbullet \ &  & \multicolumn{1}{c|}{} & \ \textbullet \ &  &  & \multicolumn{1}{c|}{} &  & \multicolumn{1}{c|}{\checkmark} &  &  &  &  &  &  &  &  &  &  &  & \multicolumn{1}{c|}{\checkmark} &  &  &  &  &  &  &  &  &  &  & \checkmark &  &  &  \\
\cite{10_arp2014drebin} & 2014 & \multicolumn{1}{l|}{Malware detection} & \ \textbullet \ &  & \multicolumn{1}{c|}{} & \ \textbullet \ &  &  & \multicolumn{1}{c|}{} & \checkmark & \multicolumn{1}{c|}{} &  &  &  &  &  &  & \checkmark &  &  &  &  & \multicolumn{1}{c|}{} &  &  &  &  &  &  &  &  &  &  &  &  &  &  \\
\cite{62_wang2016trafficav} & 2016 & \multicolumn{1}{l|}{Malware detection} & \ \textbullet \ &  & \multicolumn{1}{c|}{} & \ \textbullet \ &  &  & \multicolumn{1}{c|}{} & \checkmark & \multicolumn{1}{c|}{} &  &  &  &  &  &  &  & \checkmark &  &  &  & \multicolumn{1}{c|}{} &  &  &  &  &  &  &  &  &  &  &  &  &  &  \\
\cite{32_angelini2017goods} & 2017 & \multicolumn{1}{l|}{Malware detection} & \ \textbullet \ & \ \textbullet \ & \multicolumn{1}{c|}{} & \ \textbullet \ & \ \textbullet \ &  & \multicolumn{1}{c|}{} &  & \multicolumn{1}{c|}{\checkmark} &  &  &  &  &  & \checkmark &  &  &  &  &  & \multicolumn{1}{c|}{} &  &  &  &  &  &  & \checkmark &  &  &  & \checkmark &  &  &  \\
\cite{15_guo2018lemna} & 2018 & \multicolumn{1}{l|}{Malware detection} & \ \textbullet \ & \ \textbullet \ & \multicolumn{1}{c|}{} & \ \textbullet \ & \ \textbullet \  &  & \multicolumn{1}{c|}{} &  & \multicolumn{1}{c|}{\checkmark} &  & \checkmark &  & \checkmark &  &  &  &  &  &  &  & \multicolumn{1}{c|}{} &  &  &  &  &  &  &  &  &  &  &  &  & \checkmark &  \\
\cite{74_mathews2019explainable} & 2019 & \multicolumn{1}{l|}{Malware detection} & \ \textbullet \ & \ \textbullet \ & \multicolumn{1}{c|}{} & \ \textbullet \  & \ \textbullet \ &  & \multicolumn{1}{c|}{} &  & \multicolumn{1}{c|}{\checkmark} &  &  &  &  &  & \checkmark &  &  &  &  &  & \multicolumn{1}{c|}{} &  & \checkmark &  &  &  &  &  &  &  &  &  &  &  &  \\
\cite{27_mahdavifar2020dennes} & 2020 & \multicolumn{1}{l|}{Malware detection} & \ \textbullet \ &  & \multicolumn{1}{c|}{} & \ \textbullet \ &  &  & \multicolumn{1}{c|}{} &  & \multicolumn{1}{c|}{\checkmark} &  & \checkmark &  &  &  &  &  &  &  &  &  & \multicolumn{1}{c|}{} &  &  &  &  &  &  &  &  &  &  &  & \checkmark &  &  \\
\cite{75_kyadige2020learning} & 2020 & \multicolumn{1}{l|}{Malware detection} & \ \textbullet \ & \ \textbullet \ & \multicolumn{1}{c|}{} & \ \textbullet \  & \ \textbullet \ &  & \multicolumn{1}{c|}{} &  & \multicolumn{1}{c|}{\checkmark} &  & \checkmark &  &  &  &  &  &  &  &  &  & \multicolumn{1}{c|}{} &  & \checkmark &  &  &  &  &  &  &  &  &  &  &  &  \\
\cite{64_iadarola2021towards} & 2021 & \multicolumn{1}{l|}{Malware detection} & \ \textbullet \ &  & \multicolumn{1}{c|}{} & \ \textbullet \ &  &  & \multicolumn{1}{c|}{} &  & \multicolumn{1}{c|}{\checkmark} & \checkmark &  &  &  &  &  &  &  &  &  &  & \multicolumn{1}{c|}{} &  &  &  &  & \checkmark &  &  &  &  &  &  &  &  &  \\
\cite{65_wu2021android} & 2021 & \multicolumn{1}{l|}{Malware detection} & \ \textbullet \ &  & \multicolumn{1}{c|}{} & \ \textbullet \ &  &  & \multicolumn{1}{c|}{} &  & \multicolumn{1}{c|}{\checkmark} &  & \checkmark &  &  &  &  &  &  &  &  &  & \multicolumn{1}{c|}{} &  &  &  &  &  & \checkmark &  &  &  &  &  &  &  &  \\
\cite{76_Rabah2021iot} & 2021 & \multicolumn{1}{l|}{Malware detection} & \ \textbullet \ &  & \multicolumn{1}{c|}{} & \ \textbullet \ &  &  & \multicolumn{1}{c|}{} & \checkmark & \multicolumn{1}{c|}{\checkmark} &  & \checkmark &  &  & \checkmark & \checkmark &  & \checkmark &  &  & \checkmark & \multicolumn{1}{c|}{} &  & \checkmark &  &  &  &  & \checkmark &  &  &  &  &  &  &  \\
\cite{77_kinkead2021towards} & 2021 & \multicolumn{1}{l|}{Malware detection} & \ \textbullet \ &  & \multicolumn{1}{c|}{} & \ \textbullet \  &  &  & \multicolumn{1}{c|}{} &  & \multicolumn{1}{c|}{\checkmark} & \checkmark &  &  &  &  &  &  &  &  &  &  & \multicolumn{1}{c|}{} &  &  &  &  &  & \checkmark &  &  &  &  &  &  &  &  \\
\cite{78_koutsokostas2022microsoft} & 2022 & \multicolumn{1}{l|}{Malware detection} & \ \textbullet \ &  & \multicolumn{1}{c|}{} & \ \textbullet \  &  &  & \multicolumn{1}{c|}{} &  & \multicolumn{1}{c|}{\checkmark} &  & \checkmark &  &  &  & \checkmark & \checkmark &  &  &  &  & \multicolumn{1}{c|}{} &  &  &  &  &  &  & \checkmark &  &  &  &  &  &  &  \\
\cite{79_corona2017deltaphish} & 2017 & \multicolumn{1}{l|}{Phishing detection} & \ \textbullet \ &  & \multicolumn{1}{c|}{} & \ \textbullet \ &  &  & \multicolumn{1}{c|}{} & \checkmark & \multicolumn{1}{c|}{} &  &  &  &  &  &  & \checkmark &  &  &  &  & \multicolumn{1}{c|}{} &  &  &  &  &  &  &  &  &  &  &  &  &  &  \\
\cite{66_chai2021explainable} & 2021 & \multicolumn{1}{l|}{Phishing detection} & \ \textbullet \ &  & \multicolumn{1}{c|}{} & \ \textbullet \ &  &  & \multicolumn{1}{c|}{} &  & \multicolumn{1}{c|}{\checkmark} & \checkmark &  &  & \checkmark &  &  &  &  &  &  &  & \multicolumn{1}{c|}{} &  &  &  &  &  & \checkmark &  &  &  &  &  &  &  &  \\
\cite{67_lin2021phishpedia} & 2021 & \multicolumn{1}{l|}{Phishing detection} & \ \textbullet \ &  & \multicolumn{1}{c|}{} & \ \textbullet \ &  &  & \multicolumn{1}{c|}{} &  & \multicolumn{1}{c|}{\checkmark} & \checkmark &  &  &  &  &  &  &  &  &  &  & \multicolumn{1}{c|}{} &  &  &  &  &  &  &  &  &  &  & \checkmark &  &  &  \\
\cite{80_gulmezoglu2021xai} & 2021 & \multicolumn{1}{l|}{Privacy protection} & \ \textbullet \ &  & \multicolumn{1}{c|}{} & \ \textbullet \  &  &  & \multicolumn{1}{c|}{} &  & \multicolumn{1}{c|}{\checkmark} & \checkmark &  &  &  &  & \checkmark &  &  &  &  &  & \multicolumn{1}{c|}{} &  & \checkmark &  &  & \checkmark &  &  &  &  &  &  &  &  &  \\
\cite{26_cho2010inference} & 2010 & \multicolumn{1}{l|}{Protocol analysis} & \ \textbullet \ &  & \multicolumn{1}{c|}{} & \ \textbullet \ &  &  & \multicolumn{1}{c|}{} & \checkmark & \multicolumn{1}{c|}{} &  &  &  &  &  &  &  &  &  & \checkmark &  & \multicolumn{1}{c|}{} &  &  &  &  &  &  &  &  &  &  &  &  &  &  \\
\cite{81_fiterau2020analysis} & 2020 & \multicolumn{1}{l|}{Protocol analysis} & \ \textbullet \ &  & \multicolumn{1}{c|}{} & \ \textbullet \ &  &  & \multicolumn{1}{c|}{} & \checkmark & \multicolumn{1}{c|}{} &  &  &  &  &  &  &  &  &  & \checkmark &  & \multicolumn{1}{c|}{} &  &  &  &  &  &  &  &  &  &  &  &  &  &  \\
\cite{50_norton2017adversarial} & 2017 & \multicolumn{1}{l|}{Test-time perturbations} & \ \textbullet \ &  & \multicolumn{1}{c|}{} & \ \textbullet \ &  &  & \multicolumn{1}{c|}{} &  & \multicolumn{1}{c|}{\checkmark} & \checkmark &  &  &  &  &  &  &  &  &  &  & \multicolumn{1}{c|}{} &  &  &  &  & \checkmark &  &  &  &  &  & \checkmark &  &  &  \\
\cite{68_yamaguchi2015automatic} & 2015 & \multicolumn{1}{l|}{Vulnerability discovery} & \ \textbullet \ &  & \multicolumn{1}{c|}{} & \ \textbullet \ &  &  & \multicolumn{1}{c|}{} & \checkmark & \multicolumn{1}{c|}{} &  &  &  &  &  &  &  & \checkmark &  &  &  & \multicolumn{1}{c|}{} &  &  &  &  &  &  &  &  &  &  &  &  &  &  \\
\cite{82_russell2018automated} & 2018 & \multicolumn{1}{l|}{Vulnerability discovery} & \ \textbullet \ &  & \multicolumn{1}{c|}{} & \ \textbullet \  &  &  & \multicolumn{1}{c|}{} &  & \multicolumn{1}{c|}{\checkmark} &  &  &  &  &  & \checkmark &  &  &  &  &  & \multicolumn{1}{c|}{} &  &  &  &  &  & \checkmark &  &  &  &  &  &  &  &  \\
\cite{83_duan2019vulsniper} & 2019 & \multicolumn{1}{l|}{Vulnerability discovery} & \ \textbullet \ &  & \multicolumn{1}{c|}{} & \ \textbullet \  &  &  & \multicolumn{1}{c|}{} &  & \multicolumn{1}{c|}{\checkmark} &  & \checkmark &  &  &  &  &  &  &  &  &  & \multicolumn{1}{c|}{} &  &  &  &  &  & \checkmark &  &  &  &  &  &  &  &  \\
\cite{40_zou2020interpreting} & 2020 & \multicolumn{1}{l|}{Vulnerability discovery} & \ \textbullet \ &  & \multicolumn{1}{c|}{} & \ \textbullet \ &  &  & \multicolumn{1}{c|}{} &  & \multicolumn{1}{c|}{\checkmark} & \checkmark &  &  &  &  &  &  &  &  &  &  & \multicolumn{1}{c|}{} &  &  &  &  &  &  & \checkmark &  &  &  &  &  &  &  \\
\cite{12_alperin2020improving} & 2020 & \multicolumn{1}{l|}{Vulnerability discovery} & \ \textbullet \ &  & \multicolumn{1}{c|}{} & \ \textbullet \ &  &  & \multicolumn{1}{c|}{} &  & \multicolumn{1}{c|}{\checkmark} &  &  &  &  &  &  & \checkmark &  &  &  &  & \multicolumn{1}{c|}{\checkmark} &  & \checkmark &  &  &  &  &  &  &  &  &  &  &  &  \\
\cite{69_li2021vulnerability} & 2021 & \multicolumn{1}{l|}{Vulnerability discovery} & \ \textbullet \ &  & \multicolumn{1}{c|}{} & \ \textbullet \ &  &  & \multicolumn{1}{c|}{} &  & \multicolumn{1}{c|}{\checkmark} &  &  & \checkmark &  &  &  &  &  &  &  &  & \multicolumn{1}{c|}{} &  &  &  & \checkmark &  &  &  &  &  &  &  &  &  &  \\
\cite{36_wijekoon2021reasoning} & 2021 & \multicolumn{1}{l|}{Vulnerability discovery} & \ \textbullet \ &  & \multicolumn{1}{c|}{} & \ \textbullet \ &  &  & \multicolumn{1}{c|}{} &  & \multicolumn{1}{c|}{\checkmark} &  &  &  &  &  & \checkmark &  &  &  &  &  & \multicolumn{1}{c|}{} &  & \checkmark &  &  &  &  &  &  & \checkmark &  &  &  &  &  \\
\cite{41_de2021trust} & 2021 & \multicolumn{1}{l|}{Uncertainty estimation} & \ \textbullet \ &  & \multicolumn{1}{c|}{} & \ \textbullet \ &  &  & \multicolumn{1}{c|}{} &  & \multicolumn{1}{c|}{\checkmark} &  &  &  &  &  &  &  &  &  &  & \checkmark & \multicolumn{1}{c|}{} &  &  &  &  &  &  &  & \checkmark &  &  &  &  &  &  \\ \hline
\cite{28_chua2017neural} & 2017 & \multicolumn{1}{l|}{Binary analysis} &  & \ \textbullet \ & \multicolumn{1}{c|}{} &  & \ \textbullet \ &  & \multicolumn{1}{c|}{} &  & \multicolumn{1}{c|}{\checkmark} &  &  &  & \checkmark &  &  &  &  &  &  &  & \multicolumn{1}{c|}{} &  &  &  &  & \checkmark &  &  &  &  &  & \checkmark &  &  &  \\
\cite{61_marino2018adversarial} & 2018 & \multicolumn{1}{l|}{Intrusion detection} &  & \ \textbullet \ & \multicolumn{1}{c|}{} &  & \ \textbullet \ &  & \multicolumn{1}{c|}{} & \checkmark & \multicolumn{1}{c|}{\checkmark} &  & \checkmark &  &  &  &  & \checkmark &  &  &  &  & \multicolumn{1}{c|}{} &  &  &  &  &  &  &  &  & \checkmark &  &  &  &  &  \\
\cite{63_becker2020interpretable} & 2020 & \multicolumn{1}{l|}{Malware detection} &  & \ \textbullet \ & \multicolumn{1}{c|}{} &  & \ \textbullet \ &  & \multicolumn{1}{c|}{} &  & \multicolumn{1}{c|}{\checkmark} & \checkmark &  &  & \checkmark &  &  &  &  &  &  &  & \multicolumn{1}{c|}{} &  &  &  &  &  &  &  &  &  &  &  & \checkmark &  &  \\
\cite{17_yang2021cadeusenix} & 2021 & \multicolumn{1}{l|}{Malware detection} &  & \ \textbullet \ & \multicolumn{1}{c|}{} &  & \ \textbullet \  &  & \multicolumn{1}{c|}{} &  & \multicolumn{1}{c|}{\checkmark} &  & \checkmark &  &  &  &  &  &  &  &  &  & \multicolumn{1}{c|}{} &  &  &  &  &  &  &  &  & \checkmark &  &  &  &  &  \\
\cite{42_dolejvs2022interpretability} & 2022 & \multicolumn{1}{l|}{Malware detection} &  & \ \textbullet \ & \multicolumn{1}{c|}{} &  & \ \textbullet \ &  & \multicolumn{1}{c|}{} & \checkmark & \multicolumn{1}{c|}{\checkmark} &  & \checkmark &  &  & \checkmark & \checkmark &  &  & \checkmark &  & \checkmark & \multicolumn{1}{c|}{} &  &  &  &  &  &  &  &  &  &  &  & \checkmark &  &  \\
\cite{38_fakewake2021ccs} & 2021 & \multicolumn{1}{l|}{Privacy protection} &  & \ \textbullet \ & \multicolumn{1}{c|}{} &  & \ \textbullet \  &  & \multicolumn{1}{c|}{} &  & \multicolumn{1}{c|}{\checkmark} &  &  &  &  &  & \checkmark &  &  &  &  &  & \multicolumn{1}{c|}{} & \checkmark &  &  &  &  &  &  &  &  &  &  &  &  &  \\
\cite{37_ahn2020explaining} & 2020 & \multicolumn{1}{l|}{Traffic classification} &  & \ \textbullet \ & \multicolumn{1}{c|}{} &  & \ \textbullet \ &  & \multicolumn{1}{c|}{} &  & \multicolumn{1}{c|}{\checkmark} & \checkmark &  &  &  &  &  &  &  &  &  &  & \multicolumn{1}{c|}{} &  &  &  &  &  &  & \checkmark &  &  &  &  &  &  &  \\
\cite{84_chakraborty2021deep} & 2021 & \multicolumn{1}{l|}{Vulnerability discovery} &  & \ \textbullet \ & \multicolumn{1}{c|}{} &  & \ \textbullet \ &  & \multicolumn{1}{c|}{} &  & \multicolumn{1}{c|}{\checkmark} &  & \checkmark & \checkmark &  &  &  &  &  &  &  &  & \multicolumn{1}{c|}{} &  &  & \checkmark &  &  & \checkmark &  &  &  &  &  &  &  &  \\ \hline
\cite{70_wickramasinghe2021explainable} & 2021 & \multicolumn{1}{l|}{Anomaly detection (sensor)} &  & \ \textbullet \ & \multicolumn{1}{c|}{} &  &  & \ \textbullet \  & \multicolumn{1}{c|}{} &  & \multicolumn{1}{c|}{\checkmark} &  &  &  &  &  &  &  &  &  &  &  & \multicolumn{1}{c|}{\checkmark} &  &  &  &  &  &  & \checkmark &  &  &  &  &  &  &  \\
\cite{87_lin2020you} & 2020 & \multicolumn{1}{l|}{Backdoor injection} &  & \ \textbullet \ & \multicolumn{1}{c|}{} &  &  & \ \textbullet \ & \multicolumn{1}{c|}{} &  & \multicolumn{1}{c|}{\checkmark} & \checkmark &  &  &  &  &  &  &  &  &  &  & \multicolumn{1}{c|}{} &  & \checkmark &  &  & \checkmark &  &  &  &  &  &  &  &  &  \\
\cite{73_reyes_wifi_nids} & 2020 & \multicolumn{1}{l|}{Intrusion detection} &  & \ \textbullet \ & \multicolumn{1}{c|}{} &  &  & \ \textbullet \ & \multicolumn{1}{c|}{} & \checkmark & \multicolumn{1}{c|}{\checkmark} &  &  &  &  &  & \checkmark &  &  & \checkmark &  &  & \multicolumn{1}{c|}{} & \checkmark &  &  &  &  &  &  &  &  &  &  &  &  &  \\
\cite{85_adebayo2018sanity} & 2018 & \multicolumn{1}{l|}{Reverse engineering} &  & \ \textbullet \ & \multicolumn{1}{c|}{} &  &  & \ \textbullet \ & \multicolumn{1}{c|}{} &  & \multicolumn{1}{c|}{\checkmark} & \checkmark & \checkmark &  &  &  &  &  &  &  &  &  & \multicolumn{1}{c|}{} &  &  &  &  & \checkmark &  &  &  &  &  &  &  &  &  \\
\cite{29_alvarez2018robustness} & 2018 & \multicolumn{1}{l|}{Test-time perturbations} &  & \ \textbullet \ & \multicolumn{1}{c|}{} &  &  & \ \textbullet \ & \multicolumn{1}{c|}{} & \checkmark & \multicolumn{1}{c|}{\checkmark} & \checkmark &  &  &  &  & \checkmark & \checkmark &  &  &  &  & \multicolumn{1}{c|}{} & \checkmark & \checkmark &  &  & \checkmark &  &  &  &  &  &  &  &  &  \\
\cite{30_ghorbani2019interpretation} & 2019 & \multicolumn{1}{l|}{Test-time perturbations} &  &  & \multicolumn{1}{c|}{\ \textbullet \ } &  &  & \ \textbullet \ & &  & \multicolumn{1}{c|}{\checkmark} & \checkmark &  &  &  &  &  &  &  &  &  &  & \multicolumn{1}{c|}{} &  &  &  &  & \checkmark &  &  & \checkmark &  &  &  &  &  &  \\
\cite{86_dombrowski2019explanations} & 2019 & \multicolumn{1}{l|}{Test-time perturbations} &  &  & \multicolumn{1}{c|}{\ \textbullet \ } &  &  & \ \textbullet \ & &  & \multicolumn{1}{c|}{\checkmark} & \checkmark &  &  &  &  &  &  &  &  &  &  & \multicolumn{1}{c|}{} &  &  &  &  & \checkmark &  &  &  &  &  &  &  &  &  \\
\cite{49_zhang2020interpretable} & 2020 & \multicolumn{1}{l|}{Test-time perturbations} &  &  & \multicolumn{1}{c|}{\ \textbullet \ } &  &  & \ \textbullet \ & &  & \multicolumn{1}{c|}{\checkmark} & \checkmark &  &  &  &  &  &  &  &  &  &  & \multicolumn{1}{c|}{} &  &  &  &  & \checkmark &  &  &  &  &  &  &  &  &  \\
\cite{90_kuppa2020black} & 2020 & \multicolumn{1}{l|}{Test-time perturbations} &  &  & \multicolumn{1}{c|}{\ \textbullet \ } &  &  & \ \textbullet \ & &  & \multicolumn{1}{c|}{\checkmark} &  & \checkmark &  &  &  &  &  &  &  &  &  & \multicolumn{1}{c|}{} &  &  &  &  & \checkmark &  &  &  &  &  &  &  &  &  \\
\cite{88_galli2021reliability} & 2021 & \multicolumn{1}{l|}{Test-time perturbations} &  &  & \multicolumn{1}{c|}{\ \textbullet \ } &  &  & \ \textbullet \ & &  & \multicolumn{1}{c|}{\checkmark} & \checkmark &  &  &  &  &  &  &  &  &  &  & \multicolumn{1}{c|}{} &  &  &  &  & \checkmark &  &  &  &  &  &  &  &  &  \\
\cite{89_ghassemi2021false} & 2021 & \multicolumn{1}{l|}{Test-time perturbations} &  &  & \multicolumn{1}{c|}{\ \textbullet \ } &  &  & \ \textbullet \ & &  & \multicolumn{1}{c|}{\checkmark} & \checkmark &  &  &  &  &  &  &  &  &  &  & \multicolumn{1}{c|}{} &  & \checkmark &  &  & \checkmark &  &  &  &  &  &  &  &  &  \\ \hline
\cite{91_xu2021explainability} & 2021 & \multicolumn{1}{l|}{Backdoor injection} &  &  & \multicolumn{1}{c|}{\ \textbullet \ } &  &  &  & \multicolumn{1}{c|}{\ \textbullet \ } &  & \multicolumn{1}{c|}{\checkmark} &  &  & \checkmark &  &  &  &  &  &  &  &  & \multicolumn{1}{c|}{} &  & \checkmark &  & \checkmark &  &  &  &  &  &  &  &  &  &  \\
\cite{45_severi2021explanation} & 2021 & \multicolumn{1}{l|}{Backdoor injection} &  &  & \multicolumn{1}{c|}{\ \textbullet \ } &  &  &  & \multicolumn{1}{c|}{\ \textbullet \ } &  & \multicolumn{1}{c|}{\checkmark} &  & \checkmark &  &  &  & \checkmark & \checkmark &  &  &  &  & \multicolumn{1}{c|}{} & \checkmark &  &  &  &  &  &  &  &  &  &  &  &  &  \\
\cite{31_kuppa2021adversarial} & 2021 & \multicolumn{1}{l|}{Membership inf. \& model steal.} &  &  & \multicolumn{1}{c|}{\ \textbullet \ } &  &  &  & \multicolumn{1}{c|}{\ \textbullet \ } &  & \multicolumn{1}{c|}{\checkmark} &  & \checkmark &  &  &  &  &  &  &  &  &  & \multicolumn{1}{c|}{} &  &  &  &  &  &  &  &  & \checkmark &  &  &  &  &  \\
\cite{94_shokri2021privacy} & 2021 & Membership inference &  &  & \ \textbullet \ &  &  &  & \ \textbullet \ &  & \checkmark & \checkmark & \checkmark &  &  &  &  &  &  &  &  &  &  &  & \checkmark &  &  & \checkmark &  &  &  &  &  &  &  &  &  \\
\cite{46_zhao2021exploiting} & 2021 & \multicolumn{1}{l|}{Model inversion} &  &  & \multicolumn{1}{c|}{\ \textbullet \ } &  &  &  & \multicolumn{1}{c|}{\ \textbullet \ } &  & \multicolumn{1}{c|}{\checkmark} & \checkmark &  &  &  &  &  &  &  &  &  &  & \multicolumn{1}{c|}{} &  &  &  &  & \checkmark &  &  &  &  &  &  &  &  &  \\
\cite{92_luca2019explaining} & 2019 & \multicolumn{1}{l|}{Test-time perturbations} &  &  & \multicolumn{1}{c|}{\ \textbullet \ } &  &  &  & \multicolumn{1}{c|}{\ \textbullet \ } &  & \multicolumn{1}{c|}{\checkmark} & \checkmark &  &  &  &  &  &  &  &  &  &  & \multicolumn{1}{c|}{} &  &  &  &  & \checkmark &  &  &  &  &  &  &  &  &  \\
\cite{93_wu2021adversarial} & 2021 & \multicolumn{1}{l|}{Test-time perturbations} &  &  & \multicolumn{1}{c|}{\ \textbullet \ } &  &  &  & \multicolumn{1}{c|}{\ \textbullet \ } &  & \multicolumn{1}{c|}{\checkmark} &  & \checkmark &  &  &  &  &  &  &  &  &  & \multicolumn{1}{c|}{} &  &  &  &  & \checkmark &  &  &  &  &  &  &  &  &  \\\bottomrule
\end{tabular}

     \label{tab:full-taxonomy}
\end{table*}

\section{XAI-enabled User Assistance}\label{sec:decision-support}

The fundamental objective for employing (explainable) ML methods in security workflows is to provide decision support to model users. In fact, practitioners have been trying to make their models understandable since before the popularity of XAI, see \eg \cite{10_arp2014drebin,68_yamaguchi2015automatic,62_wang2016trafficav}. Over the past decade, numerous XAI applications have arisen to support model users in their decision-making when interacting with a deployed model. The prominence of this objective is evident from the distribution of the available literature --- 58\% of the reviewed studies provide decision support to model users.

Within the reviewed literature, the explanations are generated for distinct purposes at different levels of expertise even when considering a single stakeholder. For instance, to assist \textit{software developers} in understanding vulnerable code, some approaches simply highlight the lines of code that the model thinks are vulnerable \cite{82_russell2018automated,83_duan2019vulsniper,69_li2021vulnerability}, while others extract human-understandable rules from the vulnerable code that can serve as actionable intelligence for periodic scanning and control \cite{40_zou2020interpreting,68_yamaguchi2015automatic}. As such, these methods fall under two broad application scenarios: i) XAI is employed to provide assistance to model users for understanding model decisions and reducing their workload (\ie threat prioritization, false alarm reduction, user awareness), or ii) XAI is employed for the synthesis of new information (\ie expert knowledge creation, reverse engineering). Below we provide examples from the literature showing the different uses of explanations for different model users.

\par{\textbf{Triaging and threat prioritization.}}
\textit{Security practitioners} receive an enormous influx of cyber data that needs to be analyzed. XAI-enabled triaging techniques have been proposed to reduce analyst workload by redirecting their attention to critical events. %
This is crucial for Security Operations Center (SOC) analysts who often suffer from `alert fatigue' caused by investigating large volumes of intrusion alerts on a daily basis \cite{35_nadeem2021alert}. Black-box methods can often not be applied since the analysts are under contractual obligation to review all alerts \cite{nadeem2022learning}. Instead, XAI techniques can reduce their workload via intelligent alert management while enabling them to justify the model's decisions. 
The intuition is that a security analyst can use the explainable ML model as a `virtual assistant' that discovers meaningful patterns in large datasets and presents them to the analyst who can then make informed decisions about which data to triage and what actions to take next \cite{100_holder2021explainable}. 
For example, Nadeem \etal \cite{35_nadeem2021alert} propose alert-driven attack graphs that show attacker strategies learned from intrusion alerts. They utilize an interpretable suffix-based automaton model to learn contextually meaningful attack paths. \textit{Security analysts} can triage critical alerts by selecting one of the attack graphs.  
van Ede \etal \cite{52_robert_deepcase} use an LSTM to learn the contextual meaning of alerts by capturing the correlation between them in an attention vector. Their system clusters attention vectors, capturing attack campaigns. \textit{Security analysts} only need to analyze outlier and sampled events from emerging clusters, drastically reducing their workload. 
Similar approaches have been proposed to triage critical syslog entries for the forensic analysis of cyber attacks in a federated learning setup \cite{48_parrainterpretable2022ndss}, and to efficiently allocate cyber resources for advanced persistent threat (APT) detection \cite{39_li2021explainable}. 

\par{\textbf{False alarm reduction.}}
XAI can help \textit{security practitioners} and other \textit{model users} quickly disregard false alarms by explaining why the model made a prediction.  
For instance, Sopan \etal \cite{51_sopan2018building} propose a visual analytics dashboard to understand why an alert was raised. The dashboard provides an explanation of the alert in the form of an approximated decision path followed by the model and a list of important features. A similar approach is proposed in \cite{57_szczepanski2020achieving,59_mahbooba2021explainable}. 
Other works only show feature importance to help \textit{security analysts} understand model predictions, \eg for malware detection \cite{78_koutsokostas2022microsoft,76_Rabah2021iot,77_kinkead2021towards}, and anomaly detection \cite{13_antwarg2021explaining,47_karn2020cryptomining,56_Hwang2021esfd,71_ardito2021revisiting,72_brown2018recurrent}. In contrast, instead of explaining the predictions, de Bie \etal \cite{41_de2021trust} have proposed a metric to help \textit{security analysts} weed out false or untrustworthy predictions in regression models. They follow the intuition that instances close to each other typically have similar predictions. Thus, by comparing the prediction of a given instance with those of its k-nearest neighbours, a \textit{model user} can identify whether the prediction can be trusted.

A handful of works have used anomaly scores to automatically discard anomalous events, thus reducing the cognitive load on \textit{general model users}. For instance, Ardito \etal \cite{54_ardito2021artificial} use anomaly scores to support \textit{medical staff} in detecting when an attack has occurred on a patient's e-health telemonitoring device. The auto-encoder-based system avoids processing anomalies that can otherwise have devastating effects on a patient's health. Instead, it sends out a validation request to the medical staff. Similarly, Akerman \etal \cite{98_akerman2019vizads} use image reconstruction loss as an indication of whether artefacts in ADS-B video frames are false alarms. ADS-B is a protocol used by air traffic controllers to communicate with pilots regarding surrounding objects. By highlighting what might be false alarms, \textit{pilots} can efficiently focus on the mission at hand. 

\par{\textbf{User awareness \& education.}} XAI has been utilized to increase the general awareness of different \textit{model users} for insecure behaviour deterrence. 
For example, to keep \textit{Android users} safe, multiple works display warning signs with explanations for why an app was blocked or marked as malicious. The explanations are constructed from influential features extracted from apps' permission usage \cite{10_arp2014drebin,65_wu2021android}, and network traffic \cite{62_wang2016trafficav}.

XAI has also been used for warning \textit{end-users} when they land on potential phishing websites to improve their overall Internet browsing behaviour \cite{79_corona2017deltaphish,67_lin2021phishpedia,66_chai2021explainable}. For instance, Phishpedia \cite{67_lin2021phishpedia} employs logo detection to generate visual explanations in the form of insightful annotations on the websites. Chai \etal \cite{66_chai2021explainable} take one step further by developing a multi-modal learning setup for more accurate phishing website detection. Their attention-based explanations highlight the URL characters, website text, and images that were relevant for the detection.

Finally, in order to raise awareness among \textit{security analysts} regarding the impact of adversarial examples on a given ML model, Norton \etal \cite{50_norton2017adversarial} develop a visualization suite that lets them investigate the effect of various gradient-based adversarial attacks on image classifiers.

\par{\textbf{Expert knowledge creation.}} 
XAI can be used to synthesize human-understandable knowledge from black-box models. 
For instance, Mahdavifar \etal \cite{27_mahdavifar2020dennes} extract a surrogate rule set from a pre-trained neural network that substitutes the knowledge base of their expert system. \textit{Security analysts} interact with the expert system, which uses the rule sets to explain classification decisions. These rules are then used to classify unseen security incidents (\eg malware attacks and phishing attempts). 

In order to address the lack of interpretability in vulnerability discovery methods \cite{le2021survey}, Zou \etal \cite{40_zou2020interpreting} and Yamaguchi \etal \cite{68_yamaguchi2015automatic} extract human-understandable rules from code snippets that the model thinks are vulnerable. These rules are then used by \textit{software developers} to detect vulnerabilities in previously unseen code bases. Next to this, counterfactual explanations have been used to automatically generate patches for vulnerable code. Wijekoon \etal \cite{36_wijekoon2021reasoning} discover vulnerabilities in source code and proactively correct them with the minimal changes necessary. To this end, they use LIME to find the nearest unlike neighbour as the most similar code snippet that is not vulnerable, which is then used as a patch.

\par{\textbf{Reverse engineering.}}
Reverse engineering is commonly used in software engineering to convert black-box systems into white-box alternatives. However, there is a key difference between surrogate model learning and reverse engineering: the former extracts an interpretable model from a \textit{black-box model}, while the latter either learns an interpretable model or uses a post-hoc explainer to \textit{provide insights into the input data}. In this sense, reverse engineering methods can be considered as standalone tools that provide decision support to \textit{model users} regarding input data. 

The most common application of reverse engineering is to consider a live system as a black box, collect traces from it, and learn an interpretable model from these traces. This model can be relatively easily visualized for model-based explanations about the black-box system.
For instance, Fiterau \etal \cite{81_fiterau2020analysis} apply protocol state fuzzing on servers that use the Datagram Transport Layer Security (DTLS) protocol in order to discover functional and non-conformance issues in several implementations. Discoverer \cite{cui2007discoverer} and Prospex \cite{comparetti2009prospex} are two other popular systems for reverse engineering application-level specifications of network protocols. 
Similarly, Cho \etal \cite{26_cho2010inference} learn an automaton from botnet traffic to understand its Command and Control (C\&C) channels; 
Lin \etal \cite{55_lin2018tabor} learn an automaton from sensors of a water treatment plant to detect potential sensor malfunction, and Cao \etal \cite{34_Cao2022} learn an automaton from the network traffic of a Kubernetes cluster to identify misbehaving pods.

Alternatively, a black-box model can be learned from the traces, and post-hoc explainers can be used to explain the properties of the traces. For instance, Gulmezoglu \etal \cite{80_gulmezoglu2021xai} want to understand the type of web requests that leak side-channel information, such as performance counters and cache occupancy. This leakage enables website fingerprinting attacks in which users can be tracked by monitoring the unique combination of websites visited by their browsers. To this aim, they collect side-channel information leaked from different browsers, use it to learn several ML models, and use LIME and saliency maps to identify the leakiest web requests. 
Similarly, \textit{Malware analysts} can use reverse engineering to understand the relationship between malware samples. For instance, Nadeem \etal \cite{33_nadeem2021beyond} build behavioural profiles of malware samples by clustering their network activities. They visualize the overlap in the malware profiles in order to discover interesting malware capabilities. Similarly, Iadarola \etal \cite{64_iadarola2021towards} use gradient-based saliency maps to construct cumulative heatmaps for individual malware families that show visual differences between their disassembled code.

\subsection{The Role of Visualizations in XAI}

Visual explanations are the most common way to explain the inner workings of a black-box model. This is because human cognition prefers visual information over text for providing decision support \cite{padilla2018decision}. 
The reviewed literature proposes several types of visual explanations, \eg a graph-based interpretable automaton model proposed by Lin \etal \cite{55_lin2018tabor} that can directly be visualized for anomaly detection, and a context-based visual analytics dashboard proposed by Alperin \etal \cite{12_alperin2020improving} that uses LIME and t-SNE for triaging vulnerabilities. 

Visualizing the structure of tree-based models is another popular explanation method \cite{51_sopan2018building,57_szczepanski2020achieving,32_angelini2017goods}. Sopan \etal \cite{51_sopan2018building} report that the security analysts found their visualization of an approximated decision path generally helpful. 
However, this is not always the case. Angelini \etal \cite{32_angelini2017goods} propose a visual analytics system to explain the reason for malware detection by showing geo-locations of downloaded files and allowing a \textit{malware analyst} to drill deeper into the individual paths of a random forest. However, simultaneously exploring the paths of $\sim$100 decision trees does not make it any easier to decipher what the model is doing. Instead, it is preferable to provide different explanations based on the user's trust, \eg by providing less explanation when trust is high, and more explanation when trust is low \cite{anjomshoae2019explainable}.

Usability is an important consideration when designing decision-support tools for human analysts. Every explanation method has an associated cost in terms of its adaptation time. Even a simple XAI tool that plots reconstruction errors and lists top-k anomalies can cost analysts a full day to get used to \cite{13_antwarg2021explaining}. \textit{Generally, simpler explanations are preferred, otherwise they can make the original task even more time-consuming} \cite{panigutti2022understanding}. In other terms, complex visualizations contribute to cognitive load, subverting effective explanations. This is why the knowledge of existing analyst workflows is an important predictor in the successful deployment of XAI tools \cite{53_nyre2022considerations}. 

In addition, visualizations are not always equivalent to effective explanations. A model does not become interpretable just by virtue of visualizing it. For instance, the automaton model presented in \cite{55_lin2018tabor} requires some level of expert knowledge to correctly interpret it. 
Similarly, the decision tree proposed in \cite{59_mahbooba2021explainable} is claimed to be interpretable by default since it mimics human-level decision making, while it does not appear to be size-limited to actually be considered interpretable \cite{lipton2018mythos}.
Furthermore, the explanations provided by DeltaPhish \cite{79_corona2017deltaphish} are incomplete because they are limited to the linear coefficients of a single SVM, while it uses an ensemble of SVMs for different features.

\vspace{-7pt}
\begin{formal}
\textbf{Takeaway 1:} \textit{Visualization is not equivalent to effective explanation. XAI should reduce complexity, not add another layer of complex visualizations. }
\end{formal}
\vspace{-7pt}

\subsection{Explanation Evaluation via User Studies}

XAI-enabled user assistance tools can be evaluated along several dimensions, \eg fidelity, understandability, efficiency, and construction cost \cite{hansen2019interpretability}. Most of these criteria can be evaluated without human involvement. However, understandability involves multiple usability factors that can only be suitably evaluated with model users. This is tricky because analyst time is expensive \cite{16_han2021deepaid}. Thus, many existing works focus on evaluating other aspects of explanations instead, \eg their fidelity and efficiency \cite{70_wickramasinghe2021explainable,58_robert_frameworkwithshap,13_antwarg2021explaining}. However, an explanation is unlikely to be used in practice if it is not understandable, even if it is robust and correct. Therefore, we advise bringing humans back in the loop by evaluating XAI-enabled user assistance tools with application-grounded (with experts) or human-grounded (with lay persons) user studies \cite{doshi2017towards}.

In order to subvert costs, qualitative analyses are often conducted in place of user studies \cite{35_nadeem2021alert,17_yang2021cadeusenix,32_angelini2017goods,68_yamaguchi2015automatic,69_li2021vulnerability}. This is problematic because of the involvement of multiple stakeholders --- the XAI tools are typically developed by \textit{model designers} for \textit{model users}. In practice, these stakeholders have different expertise. We recommend avoiding qualitative analyses that only investigate the happy flows (successful explanations) in order to circumvent the possibility of cherry-picking \cite{leavitt2020towards}. 

\vspace{-7pt}
\begin{formal}
\textbf{Takeaway 2:} \textit{User studies are necessary to evaluate the usability of decision support tools. Yet, only 14\% of the reviewed literature performs user studies with a median of 8 participants.}
\end{formal}
\vspace{-7pt}

\subsection{The Importance of Stakeholder Specification}

We identified six cases within the reviewed literature where the roles of model users and designers were entangled \cite{16_han2021deepaid,15_guo2018lemna,58_robert_frameworkwithshap,32_angelini2017goods,74_mathews2019explainable,75_kyadige2020learning}. These methods assume that the same person is both, the designer and the user: 
1) Angelini \etal \cite{32_angelini2017goods} propose a visual analytics system for helping \textit{malware analysts} handle `grey cases' where a model produces a classification with low confidence. The intuition is that the explanations can either enhance the analyst's confidence in the system if the explanations make sense, or \textit{can trigger model improvement if they do not}. 
2) Kyadige \etal \cite{75_kyadige2020learning} and 3) Mathews \cite{74_mathews2019explainable} explain the output of a malware detector to help \textit{analysts} understand why a binary was classified as malicious, and \textit{evaluate whether the model uses meaningful features.}  
4) Wang \etal \cite{58_robert_frameworkwithshap} use SHAP to help \textit{security analysts} recognize the relationship between specific features and attack types, which \textit{can guide the design of a more efficient intrusion detection system}. 
5) LEMNA \cite{15_guo2018lemna} and 6) DeepAID \cite{16_han2021deepaid} are specialized XAI methods that address the unique challenges of the cybersecurity domain, \eg non-linear decision boundaries and concept drift\cite{nadeem2022intelligent}. LEMNA is a non-linear variant of LIME, while DeepAID learns a surrogate automaton model that allows users to understand the black-box model and \textit{improve it, if necessary}.

We also identified 17 cases where the intended stakeholder was left unspecified \cite{70_wickramasinghe2021explainable,71_ardito2021revisiting,72_brown2018recurrent,39_li2021explainable,73_reyes_wifi_nids,76_Rabah2021iot,77_kinkead2021towards,78_koutsokostas2022microsoft,79_corona2017deltaphish,80_gulmezoglu2021xai,38_fakewake2021ccs,28_chua2017neural,26_cho2010inference,81_fiterau2020analysis,37_ahn2020explaining,82_russell2018automated,84_chakraborty2021deep}. These methods appear to heavily focus on the fidelity of the explanations instead of their understandability, removing the human from the loop and potentially limiting their deployability. 

Disentangling stakeholders is necessary for assessing how the proposed method translates to industry, since model users and designers are often distinct parties, working in different departments or even different organizations. Moreover, since model users and designers interact with distinct phases of the ML pipeline, they often have different expertise and require disparate explanations. For example, while both Russell \etal \cite{82_russell2018automated} and Chakraborty \etal \cite{84_chakraborty2021deep} use activation to highlight vulnerable code, the former is intended for \textit{model users} (explaining why a code snippet was considered vulnerable), and the latter is intended for \textit{model designers} (making sure the model highlights correct code snippets). Even within the same domain and for the same stakeholder, the explanations can have contrasting uses, \eg within malware detection, some works use explanations to warn \textit{smartphone users} of malicious apps on their phones \cite{10_arp2014drebin,65_wu2021android,62_wang2016trafficav}, while other works provide more technical explanations to \textit{malware analysts} regarding classifier decisions \cite{78_koutsokostas2022microsoft,76_Rabah2021iot,77_kinkead2021towards}. Thus, explanations meant for one type of user might be too vague or too technical for another user \cite{blumreiter2019towards}.  

\vspace{-7pt}
\begin{formal}
\textbf{Takeaway 3:} \textit{Effective explanations are tailored to a specific user. Model users and designers usually have different expertise, and thus require disparate explanations. We encourage the community to specify their intended explanation stakeholders.
}
\end{formal}
\vspace{-7pt}

\section{XAI-enabled Model Verification}\label{sec:model-verification}

In fields other than cybersecurity, humans interact with AI systems with the assumption that they are near-perfect \cite{nicodeme2020build}. Thus, \textit{faith} or \textit{fidelity} is a major constituent of trust in the beginning, which is eventually replaced by reliance and predictability. The reverse is true for the adversarial threat landscape of cybersecurity: reliance and predictability are important constituents of trust since these systems can be attacked. To this aim, \textit{model designers} have a vital role in validating the safety and correctness of the ML pipeline in order to build trust with practitioners. 

A defensive security model designer is generally concerned with two aspects of model verification: (i) the model is robust to adversarial perturbations, and (ii) the model is generalizable and works as intended. The former is covered by the adversarial learning literature that aims to limit the possibility of evasion by making models robust to adversarial perturbations \cite{biggio2018wild,rosenberg2021adversarial}. Recent works have started to investigate the relationship between robustness and interpretability --- early evidence suggests that robust models may be more interpretable than their non-robust counterparts \cite{ross2018improving}. The intuition here is that robust models are smoother and can thus be more easily interpreted by humans. Nevertheless, further research is warranted to explore how XAI can guide the search for tamper-proof features used to train robust models.

The latter aspect of model verification fundamentally scrutinizes the generalizability of the model. Generalization is a highly desirable property in learning-based security systems as they are meant to detect previously unseen threats. To this aim, XAI has been used to detect spurious correlations --- artefacts unrelated to the security task that allow the learning algorithm to create shortcuts for separating the classes, instead of actually solving the task. These artefacts make the model \textit{seem performant} without being able to generalize in practice \cite{arp2022and,92_luca2019explaining}. In this systematization, we expand the traditional definition of spurious features to also include faulty features whose distributions are not representative of real-world cases \cite{nadeem2022intelligent}. In the literature, spurious/faulty feature detection is done via conformance checking, influential feature analysis, and surrogate model analysis. 

\par{\textbf{Conformance checking.}} Comparing classifier decisions with some notion of ground truth can be used for model debugging. For instance, 
Kyadige \etal \cite{75_kyadige2020learning} and Chua \etal \cite{28_chua2017neural} compare model outputs with expert knowledge as a sanity check to ensure that the model works correctly. Specifically, given an RNN that recovers function types and signatures from decompiled binaries, Chua \etal \cite{28_chua2017neural} use post-hoc explainers to verify that the model is able to learn concepts comparable to an expert's domain knowledge. To this aim, they use t-SNE to visualize semantically similar word embeddings, and saliency maps to understand which instructions are relevant for the recovery of the input functions. Nevertheless, many security applications struggle with obtaining ground truth, making conformance checking difficult in practice \cite{nadeem2022intelligent}.  

\par{\textbf{Influential feature analysis.}} Feature importance can be employed to investigate whether the model uses meaningful features. For example, Chakraborty \etal \cite{84_chakraborty2021deep} use LEMNA to check whether the highlighted tokens meaningfully communicate why a code snippet was classified as vulnerable. Similarly, Reyes \etal \cite{73_reyes_wifi_nids} use SHAP and Ahn \etal \cite{37_ahn2020explaining} use feature permutation to select meaningful features for intrusion detection and network traffic classification, respectively.

Feature importance can also be used to investigate causes of misclassifications in order to improve the model. For example, Becker \etal \cite{63_becker2020interpretable} propose a visual analytics system that enables \textit{malware analysts} to explore how the model views malware samples at different layers by clustering neuron activations. By visualizing the internal components of black-box models, malware analysts can identify sources of bias and misclassifications. 
Another example is from the domain of voice assistants: Chen \etal \cite{38_fakewake2021ccs} design a more robust voice assistant by first using SHAP to identify the type of fuzzy words that cause a given tree ensemble-based wake-up word detector to become falsely triggered, and then proposing countermeasures to avoid it from happening.  
Within continual learning settings, CADE \cite{17_yang2021cadeusenix} explains the cause of performance degradation of a malware detector by reporting the features that are most affected by concept drift. It uses the contrastive explanation method: it perturbs features to see which combination increases the distance to the training data the most and thus is responsible for causing drift.
Finally, Marino \etal \cite{61_marino2018adversarial} identify and correct the cause of missed detections and false alarms in IDS. They use adversarial examples that naturally serve as counterfactual explanations, showing the minimal changes required in feature values to correctly classify the (misclassified) security events. They expect that these insights will further improve their IDS performance. However, it is unclear how they avoid over-fitting since they utilize the knowledge of the test set to improve their model performance.

\par{\textbf{Surrogate model analysis.}} Interpretable surrogate models can be inferred from black-box models that can directly be inspected for defects, \eg Han \etal \cite{16_han2021deepaid} learn a surrogate automaton model, while Dolej\v{s} \etal \cite{42_dolejvs2022interpretability} learn a rule-based surrogate model. In addition, Dolej\v{s} \etal \cite{42_dolejvs2022interpretability} measure the interpretability of the surrogate model in terms of its behavioural similarity to the black-box model, \ie by checking whether they make similar mistakes. 

\subsection{The Risks of Post-hoc Explainability}

As it stands, the security literature heavily relies on performance metrics (\eg F1 score) as a means to conduct model verification. Goodhart's law dictates that when a measure becomes a target, it ceases to be a good measure \cite{clear2018atomic}. This is evident from the abundant literature on adversarial learning, suggesting that merely relying on performance metrics is a dangerous strategy as the model might have fatal weaknesses that an adversary can exploit. Moreover, high performance on experimental data does not imply that these methods would generalize in practice. This is because the analysis is rarely conducted in operational settings due to excessive costs. Furthermore, security papers often skip details on the operationalization of the ML pipeline, making it difficult to know if any spurious/faulty features have been used. As such, feature attribution can be used for identifying spurious features \cite{arp2022and}. In fact, spurious feature detection and removal should become more commonplace before deploying new models. We also recommend that publicly available models be supplemented with a verification analysis to enhance trust among practitioners. To assist \textit{model designers}, we provide an illustrative walk-through of how they can debug their models using commonplace XAI tools in \cref{sec:usecase}.

Having said that, model designers must also be aware of the risks of using post-hoc XAI for model verification: post-hoc explanations are approximations of black-box models that either hide away details or learn different concepts altogether. For example, explanations based on feature importance often disagree on the same model prediction \cite{krishna2022disagreement}, suggesting that there is a mismatch between the explanations and what the model actually does. A similar observation has been made for surrogate models \cite{Aivodji2019}. In fact, it is even possible to extract fair explanations from known unfair models. In regulated environments where companies are required to supplement their black-box models with explanations, they can be abused to perform \textit{`fairwashing'} --- promoting the false perception that a model is fair when it is actually not \cite{Aivodji2019}. Therefore, it is advisable to opt for interpretable models. Where that is not possible, it is critical to establish an equivalence relationship between a model and its explanation, \eg by learning a certifiably equivalent surrogate model. For instance, Weiss \etal \cite{weiss2018extracting} and Koul \etal \cite{koul2018learning} extract equivalent deterministic finite automata from black-box RNNs. These works fall under the safety verification literature, see \eg \cite{alshiekh2018safe,xiang2018verification,huang2020survey,gehr2018ai2}. 

\vspace{-7pt}
\begin{formal}
\textbf{Takeaway 4:} \textit{
Regardless of the XAI method used to validate a model, it is vital for safety-critical applications to establish an equivalence relation between the model and its explainer. However, this is not yet common practice within cybersecurity. 
}
\end{formal}
\vspace{-7pt}

\section{Explanation Verification \& Robustness}\label{sec:explanation-verification}

Whilst using XAI for model verification, the explanations themselves need to be verified for correctness and robustness. \textit{Model designers} are thus also responsible for conducting explanation verification to ensure the safety of the ML pipeline. This is an important line of work because XAI methods can sometimes trigger on input data patterns rather than on meaningful model behaviour. For instance, Adebayo \etal \cite{85_adebayo2018sanity} reset the weights of a neural network to their initial random values and show that some gradient-based methods still use information from the input. Therefore, evaluating the fidelity of explanations becomes vital. Yet, it has sometimes been overlooked within the security literature, see \eg \cite{57_szczepanski2020achieving,63_becker2020interpretable}. In addition, qualitative analysis alone does not provide sufficient test coverage, and may even lead to cherry-picking \cite{leavitt2020towards}. Knowing that XAI methods can generate arbitrary explanations, objective evaluation criteria are required to ensure that (a) the explanation methods work \cite{85_adebayo2018sanity,70_wickramasinghe2021explainable,87_lin2020you}, and (b) the explanations are robust to adversarial attacks \cite{43_fokkema2022attribution,29_alvarez2018robustness,chen2019robust,calzavara2020treant,44_vos2021efficient,120_hayes2022learning}. Warnecke \etal \cite{warnecke2020evaluating} and Ganz \etal \cite{ganz2021explaining} are excellent starting points for evaluation criteria for a wide variety of post-hoc explainers under security settings. Their criteria include descriptive accuracy, sparsity, completeness, stability, efficiency, and robustness.

\par{\textbf{Fidelity evaluation.}} The explanation fidelity can be evaluated in several ways: Wickramasinghe \etal \cite{70_wickramasinghe2021explainable} test the fidelity of their attribution method by perturbing feature values and analyzing their impact on the explanations. Lin \etal \cite{87_lin2020you} test the correctness of different saliency explanations by deliberately injecting artefacts in the input data to see if the explanations detect them. Specifically, they inject backdoor trigger patterns in input images that would naturally result in misclassifications by a CNN. These backdoor triggers serve as ground truth, \ie the backdoor features are primarily responsible for causing misclassifications, so a faithful explainer must be able to identify them.

\par{\textbf {Adversarial robustness.}} More importantly, XAI forms an additional attack vector for \textit{adversaries} within the context of cybersecurity --- both post-hoc explainers \cite{30_ghorbani2019interpretation,86_dombrowski2019explanations,90_kuppa2020black,49_zhang2020interpretable,88_galli2021reliability,89_ghassemi2021false} and interpretable models \cite{biggio2013evasion,lowd2005adversarial} are sensitive to small adversarial perturbations.
For instance, Ghorbani \etal \cite{30_ghorbani2019interpretation} investigate the effect of adversarial perturbations on exemplars. Exemplars are samples from the training set whose features most resemble the instance to be explained. They find that while keeping the prediction equal, they can cause the top-3 exemplars to be entirely different for perturbed samples, implying that the perturbed samples enter a part of the model with drastically different latent features.
Moreover, Dombrowski \etal\cite{86_dombrowski2019explanations} exploit the fragility of explanations to perform targeted attacks. They show that by adding imperceptible perturbations to the input image, the adversary can completely control the generated explanation.
These studies identify three problematic traits of post-hoc explainers:
\begin{itemize}
    \item Predictions and explanations can change tremendously under small perturbations;
    \item While keeping the explanation fixed, input samples can be perturbed to cause misclassifications;
    \item While keeping the prediction fixed, input samples can be perturbed to change the explanation.
\end{itemize}

The fact that models and explainers can be attacked independently opens up a new range of attacks. For instance, malware authors can evade detection while masking the features that they used for evasion. In this case, the generated feature importance maps do not represent the features that are actually important for classification. Therefore, it is imperative that \textit{model designers} robustify explainers against adversarial perturbations. 

Recent works have started to investigate the robustness of post-hoc explainers: Alvarez-Melis \etal \cite{29_alvarez2018robustness} investigate the smoothness of explanations around data points as a measure of robustness. Based on a local version of the Lipschitz constant, they show that the smoothness of model-agnostic explainers, \eg LIME and SHAP, can vary across datasets. They also show that gradient-based explanations are approximately four times smoother than LIME, suggesting that model-based explanations are more robust than their model-agnostic counterparts. For counterfactual explanations, Fokkema \etal \cite{43_fokkema2022attribution} show that robust explainers cannot also be recourse sensitive\footnote{Recourse refers to a description of feature modifications required to change the model outcome.}. This means that there will always be model inputs for which the explanations suggest modifications 
that do not end up changing the model's prediction\footnote{Note that the inputs for which this happens might not occur in practice and that this problem does not exist in linear unbounded models.}. As a solution, they suggest using multiple counterfactual explanations pointing in different directions.

A handful of works have also proposed robust variants of interpretable models, such as linear models and decision trees. For instance, Vos \etal\cite{44_vos2021efficient} learn efficient and robust decision trees, while Hayes \etal\cite{120_hayes2022learning} learn robust and differentially private logistic regression. Note that decision trees and logistic regression are considered interpretable as long as they are size-limited \cite{lipton2018mythos}.

\vspace{-7pt}
\begin{formal}
\textbf{Takeaway 5:} \textit{Along with ML models, explainers can also be attacked. In addition to fidelity testing, we recommend either using a robust explainer or conducting explanation verification under adversarial settings.}
\end{formal}
\vspace{-7pt}

\section{Offensive Use of Explanations}\label{sec:offensive-xai}

From the offensive security perspective, XAI can also provide decision support to \textit{adversaries} for better attack formulation. As outlined by Papernot \etal \cite{papernot2018sok}, adversaries can target multiple phases of an ML pipeline, \eg the training phase for poisoning attacks, and the deployment phase for evasion and privacy attacks. XAI can further strengthen these capabilities by exposing sensitive details about the model. 
Adjusting the definitions proposed in \cite{papernot2018sok} for XAI, we organize the nefarious uses of explainers through the lens of the classical confidentiality, integrity, and availability (CIA) triad \cite{pfleeger2012analyzing}. Considering the added utility of XAI, attacks on \textit{confidentiality} utilize explanations to expose the model structure or the data on which the model was trained.
Attacks on \textit{integrity} and \textit{availability} utilize explanations to discover knowledge that adversaries can use to induce specific model outcomes of the adversary's choosing and thwart legitimate users from accessing meaningful model outputs. 

\par{\textbf{Confidentiality attacks.}}
Explanations provide additional information to \textit{adversaries} about the inner workings of a deployed model, making it easier to reconstruct the model and the training data. This is why explanations are seen as privacy vulnerabilities \cite{hall2019}. Yet, little work is done to generate privacy-preserving explanations \cite{biswal2022system}.
In the literature, XAI has been used to strengthen model inversion \cite{46_zhao2021exploiting}, membership inference \cite{94_shokri2021privacy} and model extraction attacks \cite{31_kuppa2021adversarial}. 

Model inversion attacks enable adversaries to reconstruct training data from model predictions \cite{papernot2018sok}. Adversaries can reproduce the model more accurately by utilizing explanations, \eg Zhao \etal \cite{46_zhao2021exploiting} use an XAI-aware model inversion attack to successfully recover images from the training data. They show that feature importance maps generated from gradients and layer-wise relevance propagation (LRP) helped improve image reconstruction and led to an increase in model inversion performance compared to only using predicted probabilities.

Membership inference attacks assume that an adversary has some inputs and they want to predict whether they were used during training \cite{papernot2018sok}. Shokri \etal \cite{94_shokri2021privacy} utilize gradient-based explanations to perform stronger membership inference attacks. They use the variance of saliency maps as a feature to infer membership and show that it works better than mere random guessing. The performance further improves when using the full explanation instead of only the variance.

Finally, model extraction attacks enable adversaries to recover the model's structure and parameters from predictions \cite{papernot2018sok}. Kuppa \etal \cite{31_kuppa2021adversarial} utilize counterfactual explanations to improve their model extraction and membership inference attacks. They perform the model extraction attack by learning a surrogate model from known predictions and explanations. In addition, they perform membership inference by comparing the predictions of the target and counterfactual models to infer whether an input belonged to the training data.
 
\par{\textbf{Integrity and Availability attacks.}}
Explanations provide additional knowledge to \textit{adversaries} about the features to perturb in order to alter the correct functioning of the model. This can be done while the model is already deployed (\ie evasion attacks) or when the model is training (\ie poisoning and backdoor attacks). 

Demetrio \etal \cite{92_luca2019explaining} use integrated gradients to explain the importance assigned to the various fields of binary executables. They use this information to identify a few bytes in the malware header that need to be perturbed in order to successfully evade detection. 

Poisoning attacks are specialized adversarial attacks where an adversary injects a small percentage of perturbed data to get some desired change in the learned model. 
Kuppa \etal \cite{31_kuppa2021adversarial} use counterfactual explanations to find the malware features that most heavily impact the classifier decision. They use this knowledge to craft adversarial training samples that efficiently poison the model.

Backdoor attacks are specialized poisoning attacks where the adversary makes the model sensitive to a pre-specified trigger. 
Severi \etal \cite{45_severi2021explanation} use SHAP to craft backdoor triggers in malware detectors. Utilizing the explanation, they determine which features to poison, resulting in a success rate of up to three times higher than that of a greedy algorithm that does not use XAI.
Similarly, Xu \etal \cite{91_xu2021explainability} inject backdoors into GNNs by leveraging XAI techniques. They employ GNNExplainer to identify the parts of the graph to attack, and GraphLIME to identify the node features and values to change.

In two-player competitive games, Wu \etal \cite{93_wu2021adversarial} utilize XAI to exploit the weakness of an adversarial reinforcement learning agent. In such games, the agents take optimal actions according to their policy function, which is often learned using self-play. Using saliency maps, the proposed adversarial agent observes which of their actions the opponent pays the most attention to, and alters them in the next time stamp, thus confusing and manipulating the opponent's actions.

\vspace{-7pt}
\begin{formal}
\textbf{Takeaway 6:} \textit{Uniquely attributed to the security domain, adversaries may abuse explanations to bolster their capabilities. Meanwhile, research on privacy-preserving explanations that are also robust to evasion attacks is almost non-existent.}
\end{formal}
\vspace{-7pt}

\section{Tutorial: Debugging a Malicious Network Traffic Detector via XAI}\label{sec:usecase}

Sections \cref{sec:model-verification} and \cref{sec:explanation-verification} elucidate the critical role of model designers in ensuring the correctness and robustness of an ML pipeline. While XAI has been commonly directed towards model users, we argue that model designers can also greatly benefit from it. 
In this section, we present an illustrative tutorial on how model designers might use XAI for model verification. 
Specifically, we demonstrate how the investigation of influential features and misclassifications can identify problematic or spurious features. The experiments necessitate a sufficient understanding of post-hoc explainers for correct interpretation and highlight the expressive power of interpretable models.
 
We consider a \textit{model designer} who learns an ML model to detect malicious botnet traffic on their company's network. They have some intuition of how a potential botnet-infected device might behave, and thus use XAI to validate whether the model follows that intuition, \eg by checking whether it uses any spurious features or any strange artefacts from the training data. 
Note that we are only interested in finding spurious features: while the selection of tamper-resistant features is also an important problem, using XAI to discover such features remains an open problem, to the best of our knowledge.

\par{\textbf{Dataset selection.}} We use the open-source CTU-13~\cite{ctu_13} as our experimental dataset. It has 13 scenarios, each containing both benign and malicious Netflow data. The malicious Netflows in each scenario are collected by monitoring virtual machines (VM) infected with real malware. Each Netflow has the following features: start time (StartTime), duration (Dur), protocol (Proto), source port (Sport), Netflow direction (Dir), destination port (Dport), state (State), source type of service (sTos), destination type of service (dTos), total packets (TotPkts), total bytes (TotBytes), and source bytes (SrcBytes). The dataset contains 64,855,215 benign and 1,535,374 malicious Netflows.

\par{\textbf{Experimental setup.}} We have developed a modular XAI pipeline in Python with six models and four explainers. Implementation details are given in appendix \ref{pipeline-design}. We release the code for reproducibility\footnote{
XAI pipeline: \url{https://github.com/tudelft-cda-lab/xai-pipeline}
}.

For the experiments, we train a gradient boosting machine (GBM) over all the features of the Netflow data, as described in \cite{ctu_13}.  The GBM achieves a balanced accuracy of 86.4\%. While this model is arguably not state-of-the-art for detection purposes, it is a black box that concretely shows how improvements can be obtained via XAI. As such, the analysis described in this section can be applied to any black-box model\footnote{We recognize that the tutorial discusses a simple case study and that the features may have more complex relationships in reality. However, even this simple case occurs frequently in practice, as shown in \cite{d2022establishing}.}.

\begin{figure}[t]
    \centering
    \includegraphics[width=\linewidth]{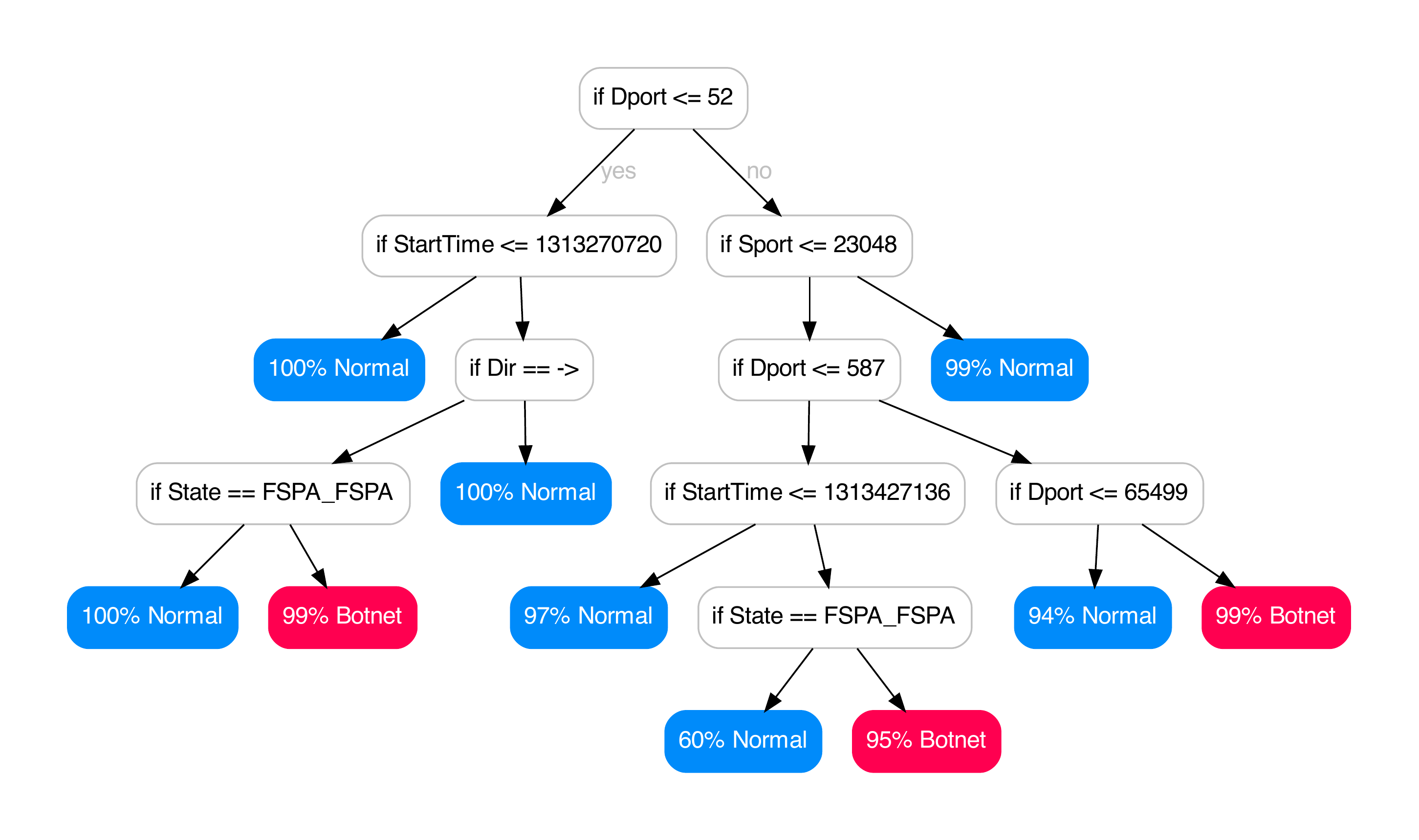}
    \caption{Decision tree for the CTU-13 dataset. It uses StartTime and Sport to differentiate between benign and malicious Netflows.}
    \label{fig:decision-tree}
\end{figure}

We use SHAP, LIME and LEMNA to explain the predictions of the GBM. We also learn an interpretable decision tree (see Figure \ref{fig:decision-tree}) to verify whether similar conclusions can be drawn from model-based and model-agnostic explanations. The decision tree has nine nodes and achieves a balanced accuracy of 83.6\%, which is only slightly worse than the GBM. The SHAP summary plot and LEMNA explanations\footnote{LEMNA is excluded from the analysis since it provides remarkably fewer insights for class distinction compared to SHAP and LIME.} for the GBM are in appendix \ref{extra-explanations}. 
We generate explanations for 140 Netflows from the test set: 50 true positives (malicious), 50 true negatives (benign), 20 false positives (not malicious), and 20 false negatives (not benign). 

\begin{figure*}[t]
    \centering
    \includegraphics[width=0.19\linewidth]{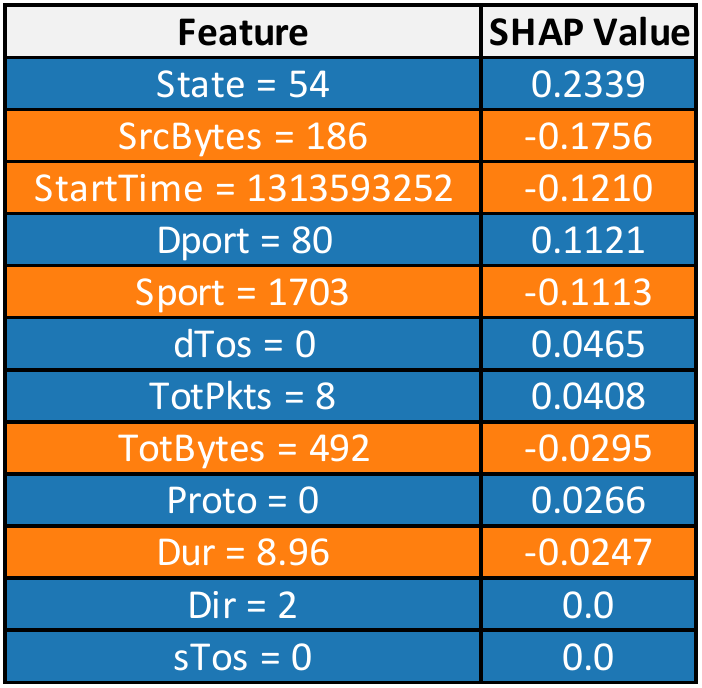}
    \includegraphics[width=0.35\linewidth]{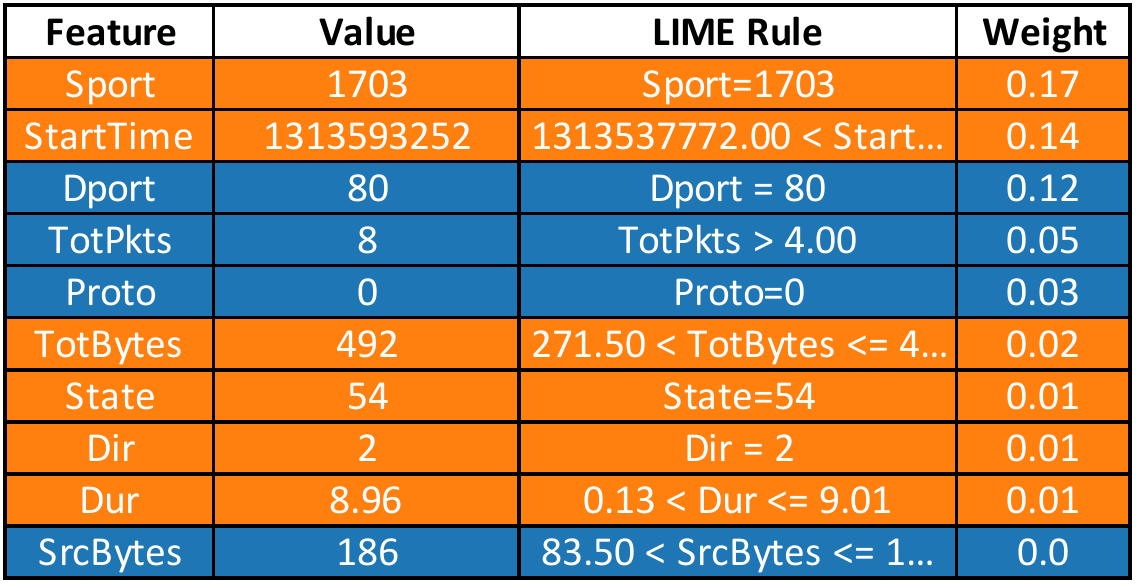}
        \includegraphics[width=0.2\linewidth]{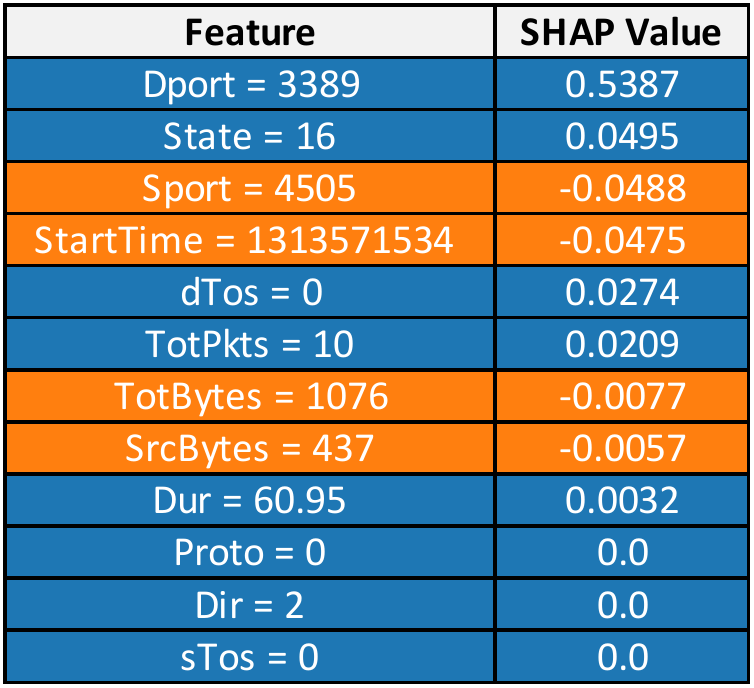}
        \caption{(Left): SHAP explanation for a false positive Netflow. (Middle): LIME explanation for a false positive Netflow. (Right): SHAP explanation for a false negative Netflow. Orange rows contribute positively, and blue rows contribute negatively towards the malicious label.}
    \label{fig:shap-lime}
\end{figure*}

\par{\textbf{1. XAI for discovering spurious correlations.}} It is evident from the SHAP summary plot (Figure \ref{fig:global-shap}) that the GBM exhibits a strong reliance on the destination port, source port, and start time features. 
We also see this trend in the interpretable decision tree in Figure \ref{fig:decision-tree}. 

The reliance on the start time and source port features is problematic: start time is problematic because it represents Unix time, so each new Netflow will have a vastly different feature value compared to the ones seen in the training data, negatively impacting the test accuracy and the model's generalizability. For instance, a benign Netflow that the GBM considers as malicious with a probability of 65\% suddenly becomes benign with a probability of 94\% if we artificially perturb the start time to four weeks earlier. This implies that the model learns to predict \textit{when a Netflow is generated}, rather than the Netflow's maliciousness. 

Source port is problematic because it typically gets arbitrarily assigned by the operating system, and as such should not be indicative of malicious behaviour. However, the CTU-13 dataset uses only a small subset of VM-related port numbers \cite{cao2022encoding}, which inadvertently becomes indicative of malicious behaviour. This is a common shortcoming of lab-collected datasets \cite{arp2022and}. Thus, it can also be considered an artefact of the experimental data.

It is noteworthy that start time and source port are perfectly valid features if the test set comes from CTU-13. Since we cannot expect real data to follow the same patterns as CTU-13, we consider them spurious features. This type of analysis is not common practice in the security literature: several recent and relatively popular works utilize the identified faulty features, see \eg \cite{nicolau2018learning,khan2019adaptive,priyadarshini2019deep}. Since these features are tightly coupled with the prediction label, standard feature selection methods are unlikely to get rid of them. This is where XAI can help. 

The next logical step is to retrain the model without the spurious features. Doing so lowers the balanced accuracy of the GBM and decision tree to 74.4\% and 58.1\%, respectively. We argue that this is an improvement since the faulty features were making the classifier \textit{appear performant} without being able to generalize in practice. 
 Because cyber data is often noisy, sole reliance on performance metrics is generally meaningless, especially when spurious features are involved. Therefore, we recommend that like ablation studies, the identification and removal of spurious features should become a fundamental step in the design of ML pipelines.

\par{\textbf{2. XAI for finding causes of misclassifications.}}
We find that post-hoc explanations must be supplemented with input data statistics to make meaningful inferences regarding the causes of misclassifications. 
For instance, we analyze a randomly sampled false positive Netflow. The local SHAP explanation (see Figure \ref{fig:shap-lime}a) shows a heavy reliance on the state value of 54 and source bytes of 186. This information in itself is likely insufficient for an analyst to understand why the model made this mistake. However, combining this information with an analysis of the training data reveals that these feature values appear almost exclusively in malicious samples, thus identifying the cause of the false positive. 

In another example, we analyze a randomly sampled false negative Netflow. The local SHAP explanation (see Figure \ref{fig:shap-lime}c) shows a substantial reliance on the destination port 3389, which is associated with the remote desktop protocol (RDP). Internet-facing RDP servers commonly fall victim to cyber attacks\footnote{\href{http://darktrace.com/blog/botnet-malware-remote-desktop-protocol-rdp-attack}{http://darktrace.com/botnet-malware-remote-desktop-protocol-rdp}}, making it a likely indicator of suspicious activity. Yet strangely, the port 3389 has contributed heavily towards the opposite. 
Analyzing the training data reveals that RDP is mostly used by benign hosts in the CTU-13 dataset, due to which the model incorrectly classifies a malicious Netflow as benign. These examples reveal what appear to be sampling and confounding biases in the CTU-13 dataset. 

\vspace{-7pt}
\begin{formal}
\textbf{Takeaway 7:} \textit{Feature importance explanations do not provide the full picture in isolation. Instead, actionable insights can be obtained by combining the input data together with post-hoc explanations.}
\end{formal}
\vspace{-7pt}

\par{\textbf{3. Utility of different XAI types.}} 
All explanations are not created equal. Since XAI is meant to explain the behaviour of a model, testing the predictability of the model on a new (previously unseen) data instance, given a few explanations provides a simple estimate of the explanation's utility. In this sense, there is a clear divide between interpretable models and post-hoc explanations.

For a given interpretable model, such as the decision tree in Figure \ref{fig:decision-tree}, it is almost trivial to predict how a new instance will be classified by following the decision path.
However, since post-hoc explanations are mere approximations of the black-box model, it is difficult to predict how the GBM would classify a new instance, given its LIME and SHAP explanations. For instance, the local SHAP explanations provide feature importance with equality relationships (\eg see Figure \ref{fig:shap-lime}a), which makes it impossible to predict how a new instance will be classified, even if it resembles the instances for which explanations are already available. This is because the explanations do not reveal the impact of slight feature perturbations on the classification. We encountered almost the same problem for LIME even though it considers a local neighbourhood to prevent this very issue. 

Moreover, post-hoc explainers compute their local neighbourhoods differently, causing explanations for the same model prediction to differ. Going back to the false positive example, SHAP (Figure \ref{fig:shap-lime}a) heavily relies on the state feature, while LIME (Figure \ref{fig:shap-lime}b) assigns very low importance to it. Also, while SHAP considers dTos to be important, it does not even appear in LIME. This disagreement problem between feature attribution methods has recently been discussed by Krishna \etal \cite{krishna2022disagreement}.
Based on their metric, there is a 25.5\% disagreement rate between the top-3 features of SHAP and LIME for our 140 Netflows.
This exemplifies the mismatch between the black-box model and its explanations and makes a strong case for learning interpretable models from the get-go.

Furthermore, the correct interpretation of post-hoc explanations often relies on how well the explainee understands the underlying mechanisms of the method, reiterating the importance of user studies in explanation evaluation. For instance, while both local-SHAP and LIME show feature importance, their explanation interpretation can be very different. We found that LIME assigns very low weights to all features for almost all the Netflows. This does not imply that similar Netflows should have the same label, as one would intuitively expect, but rather that LIME has low confidence about the prediction given its local surroundings. Thus, an unsuspecting analyst might draw misleading conclusions by overly relying on intuition rather than the understanding of the method \cite{leavitt2020towards}.

\vspace{-7pt}
\begin{formal}
\textbf{Takeaway 8:} \textit{LIME and local-SHAP are both feature attribution methods but their interpretations can be different. Working knowledge of post-hoc explainers is cardinal for correctly interpreting the explanations.}
\end{formal}
\vspace{-7pt}

\section{Discussion and Open Problems}\label{sec:discussion}

Below, we identify open problems and provide recommendations for further XAI research within cybersecurity:

\noindent\par{\textbf{User study crisis.}} The lack of qualitative validation among decision support papers is alarming. While \textit{model users} are the most common consumers of explanations in the security literature, they have regularly been excluded from the evaluation process \textit{(Takeaways 1-3, 8)}. The evaluation of robustness and fidelity does not guarantee usability, which is arguably an equally important trait of good explanations. However, usability is rarely taken into account when designing evaluation criteria for effective security explanations, see \eg \cite{warnecke2020evaluating,ganz2021explaining}.
Since analyst time is expensive, it may be beneficial to develop proxy tasks and metrics on which to evaluate new research instead. A handful of studies have incorporated human cognition in their metric definition. For example, Islam \etal \cite{islam2020towards} quantify the complexity of post-hoc explanations in terms of cognitive chunks, and Dolej\v{s} \etal \cite{42_dolejvs2022interpretability} quantify the added opaqueness of explanations \textit{w.r.t.} known interpretable models. 
Alternatively, in the absence of security practitioners, newly developed tools could be peer-reviewed regarding their usability, \eg during conference artefact evaluation sessions. 
Furthermore, disentangling and specifying stakeholders should also provide clarity regarding the intended subjects for user studies. 

\noindent\par{\textbf{Robustness \textit{vs.} interpretability.}} The role of \textit{model designers} is minimized in the security literature with merely 22.3\% of the literature focused on model \& explainer verification \textit{(Takeaways 4-5, 7-8)}. 
Since trust manifests inherently differently in the security domain, specialized XAI methods are needed to bolster practitioner trust in ML pipelines. 
While the tutorial in \cref{sec:usecase} helps model designers get started with XAI-enabled model verification, the role of XAI in tamper-resistant feature selection and robust model learning remains unclear. 
Another related question is regarding the relationship between robustness and interpretability: initial research eludes to robust models being more interpretable than non-robust models \cite{ross2018improving}, requiring further research in this direction.

\noindent\par{\textbf{Price of interpretability.}} If interpretable surrogate models are to be used for model verification, they must be certifiably equivalent to their black-box parent models for the evaluation to be meaningful. In this sense, directly learning a robust interpretable model (as opposed to learning a black-box model explained by a surrogate) may prove more helpful in establishing trust. Yet, only 25.9\% of the studies we reviewed adopt interpretable models, while the majority of them focus on applying post-hoc explainability. The discussion regarding the \textit{`price of interpretability'} (measuring the trade-off between explainability and performance) \cite{bertsimas2019price} requires special considerations in cybersecurity. We believe that the presence of an adversary and the prevalence of spurious features will likely make this trade-off less pronounced compared to other fields. However, further research is warranted in this area.
Furthermore, it may be possible to exploit the power of post-hoc explanations without losing interpretability: black-box models may be used as a benchmark to guide the search for better interpretable models. Post-hoc explanations can provide actionable intelligence regarding features and parameters for interpretable models. In this way, we view post-hoc explainers and interpretable models as complementary methods rather than as alternatives.

\noindent\par{\textbf{Privacy-preserving explanations.}}
In addition to attacking the XAI module, adversaries can also utilize explanations, much like \textit{model users}, but with a devious intent \textit{(Takeaway 6)}. 
This makes it difficult to provide explanations to model designers and model users without the adversaries also taking advantage of them. There is some preliminary work that studies the trade-off between explainability and privacy in order to select privacy-preserving explanations \cite{biswal2022system}. However, such explanations could still be used to bolster attacks on model integrity and availability. Therefore, this is also an urgent avenue for future research.

\section{Conclusions}
We systematize available research that utilizes explainable models for solving security problems. We identify 3 cybersecurity stakeholders that employ XAI for 4 research objectives within a typical ML pipeline. 
Distilled from a diverse body of literature, this overview streamlines existing research on explainability within cybersecurity and provides a starting point for practitioners.

We found evidence that the security literature does not always disentangle model users and designers.
In addition, only 22.3\% of the security literature focuses on model \& explanation verification. 
This is problematic because model designers have a critical role in ensuring the correctness and security of an ML pipeline. With regards to model correctness, we specifically provide a walk-through tutorial of how model designers can successfully detect and discard spurious features using SHAP \& LIME. At the same time, the example also exposed the disagreement problem between local explanations and showed that SHAP \& LIME have different interpretations. Thus, model designers must have a working knowledge of the explanation method in order to draw correct conclusions.

Moreover, adversaries can not only attack the XAI component but can also utilize explanations to compromise the confidentiality, integrity and availability of a model. Meanwhile, research on limiting these abuses is almost non-existent. Finally, the lack of user validation in XAI-enabled user assistance, and the lack of interpretability by design shows the substantial margins of improvement within the field of XAI for cybersecurity. 

\subsubsection*{Acknowledgements} We thank Dani\"el Meinsma for his contributions to the literature review, and the anonymous reviewers for their valuable feedback. This work was made possible by TTW VIDI project 17541 (LIMIT) and EU H2020 project 952647 (AssureMOSS). 

\bibliographystyle{unsrt}
\bibliography{main}

\appendices

\section{Selected Literature}\label{selected-lit}

\begin{figure}[t]
    \centering
    \includegraphics[width=0.9\linewidth]{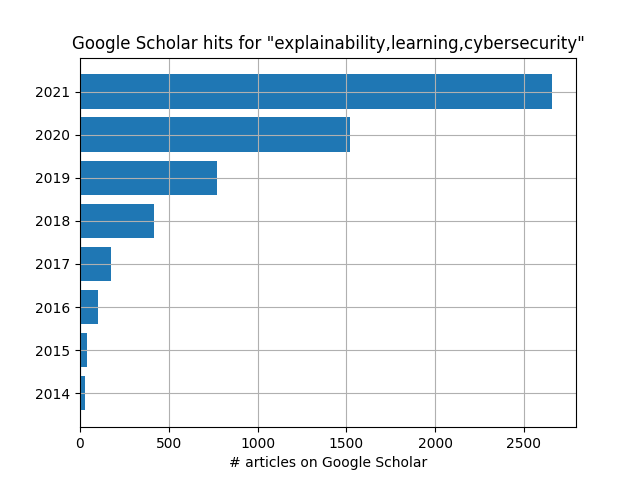}
    \caption{Prevalence of XAI literature from 2014-2021.}
    \label{fig:lit-prev}
\end{figure}

The literature related to \textit{explainability} and \textit{cybersecurity} has increased dramatically since 2014, see Figure \ref{fig:lit-prev}. This literature is fragmented across various Computer Science domains. Table \ref{tab:venues} provides a list of venues from where the reviewed literature was selected. 

\begin{table}[t]
\caption{Venues investigated for literature discovery}
\label{tab:venues}
\rowcolors{2}{lightgray}{}
\centering
\resizebox{\columnwidth}{!}{%
\setlength{\tabcolsep}{2pt}
\tiny
\begin{tabular}{llc}
\toprule
\textbf{Domain} & \textbf{Type} & \textbf{Venue} \\ \midrule
Cybersecurity & Conference & ACM   Conference on Computer and Communications Security (CCS) \\ 
Cybersecurity & Conference & Asia Conference on Computer and Communications Security   (AsiaCCS) \\ 
Cybersecurity & Conference & European Symposium on Research in Computer Security (ESORICS) \\ 
Cybersecurity & Conference & European Symposium on Security and Privacy (Euro S\&P) \\ 
Cybersecurity & Conference & IEEE   Symposium on Security and Privacy (S\&P) \\ 
Cybersecurity & Conference & International Conference on Security and Privacy \\ 
Cybersecurity & Conference & Italian Conference on Cybersecurity (ITASEC) \\ 
Cybersecurity & Conference & Network   and Distributed System Security (NDSS) \\ 
Cybersecurity & Conference & USENIX   Security Symposium \\ 
Cybersecurity & Journal & Computers and Security \\ 
Cybersecurity & Journal & IEEE Transactions on Dependable and Secure Computing (TDSC) \\ 
Cybersecurity & Journal & IEEE   Transactions on Information Forensics and Security (TIFS) \\ 
Cybersecurity & Journal & IEEE Transactions on Networks and Systems Management (TNSM) \\ 
Cybersecurity & Workshop & ACM workshop   on Artificial Intelligence and Security (AISec) @ CCS \\ 
Cybersecurity & Workshop & ACM workshop on Wireless Security and Machine Learning \\ 
Cybersecurity & Workshop & \begin{tabular}[c]{@{}c@{}}AI-enabled Cybersecurity   Analytics and Deployable Defense (AI4Cyber) \end{tabular} \\ 
Cybersecurity & Workshop & IEEE Symposium on Visualization for Cybersecurity (VizSec) @   VIS \\ 
Cybersecurity & Workshop & Machine   Learning for Cyber Security (MLCS) @ ECML/PKDD \\ 
Cybersecurity & Workshop & Workshop on Artificial Intelligence and Cybersecurity   (AI-Cybersec) \\ 
Cybersecurity & Book & Malware Analysis using Artificial Intelligence and Deep   learning (Springer) \\ \hline
Machine learn. & Conference & AAAI Conference on Artificial Intelligence \\ 
Machine learn. & Conference & ACM Conference on Knowledge Discovery \& Data Mining   (SIGKDD) \\ 
Machine learn. & Conference & Conference   on Neural Information Processing Systems (NeurIPS) \\ 
Machine learn. & Conference & International Conference on Computer Vision \\ 
Machine learn. & Conference & International Conference on Intelligence Virtual Agents \\ 
Machine learn. & Conference & International Conference on Machine Learning (ICML) \\ 
Machine learn. & Conference & International Conference on Pattern Recognition \\ 
Machine learn. & Conference & International   Joint Conference on Artificial Intelligence (IJCAI) \\ 
Machine learn. & Conference & International Joint Conference on Neural Networks (IJCNN) \\ 
Machine learn. & Journal & Advances in Intelligent Systems and Computing (Springer) \\ 
Machine learn. & Journal & Human-Intelligent Systems Integration (Springer) \\ 
Machine learn. & Journal & Nature Machine Learning \\ 
Machine learn. & Journal & Neural Computing and Applications (Springer) \\ \hline
Computer Sci. & Conference & \begin{tabular}[c]{@{}c@{}}ACM Symposium on   High-Performance \\ Parallel and Distributed Computing (HPDC)\end{tabular} \\ 
Computer Sci. & Conference & Conference on   Human Factors in Computing Systems (CHI) \\ 
Computer Sci. & Conference & International Conference on Enabling Technologies (WETICE) \\ 
Computer Sci. & Conference & International Conference on Human System Interactions (HSI) \\ 
Computer Sci. & Journal & ACM Computing Surveys \\ 
Computer Sci. & Journal & Annual Conference on Industrial Electronics Society (IECON) \\ 
Computer Sci. & Journal & Electronics (MDPI) \\ 
Computer Sci. & Journal & Expert Systems with Applications (Science Direct) \\ 
Computer Sci. & Journal & IEEE Access \\ 
Computer Sci. & Journal & Lancet Digital Health (Elsevier) \\ 
Computer Sci. & Journal & Procedia Computer Science (Science Direct) \\ 
Computer Sci. & Journal & Quality and Reliability Engineering (Wiley) \\ \hline
Software Engg. & Conference & \begin{tabular}[c]{@{}c@{}}ACM Joint European Software Engineering Conference and \\ Symposium on Foundations of Software Engineering (ESEC/FSE)\end{tabular} \\ 
Software Engg. & Conference & \begin{tabular}[c]{@{}c@{}}ACM/IEEE International Conference on Model Driven \\ Engineering   Languages and Systems (MODELS)\end{tabular} \\ 
Software Engg. & Journal & ACM Transactions on Software Engineering \\ 
Software Engg. & Journal & ACM Transactions on Software Engineering and Methodology \\ 
Software Engg. & Workshop & \begin{tabular}[c]{@{}c@{}}International workshop on Continuous Software \\ Evaluation and   Certification (IWCSE) @ ARES\end{tabular} \\ \bottomrule
\end{tabular}}
\end{table}

\begin{figure}[t]
    \centering
    \includegraphics[width=\linewidth]{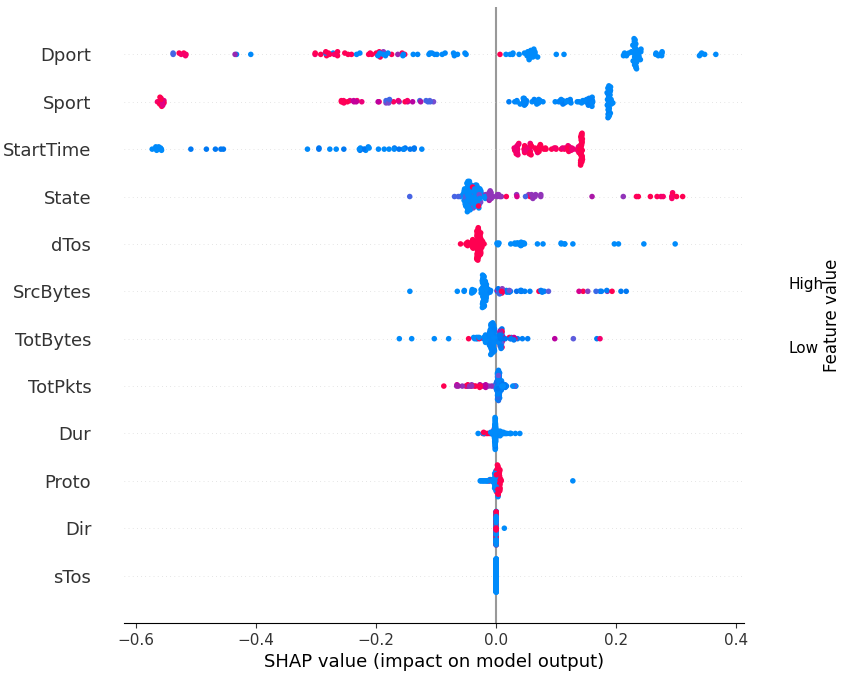}
    \caption{Global SHAP summary plot for the GBM. The y-axis shows the features ordered by their importance. For each feature, the Netflows are depicted as points, and their colour represents the contribution of the specific feature towards the model output.}
    \label{fig:global-shap}
\end{figure}

\begin{figure*}[t]
    \centering
    \includegraphics[width=0.21\linewidth]{Figures/experiment/shap_139_fp_table.pdf}
    \includegraphics[width=0.39\linewidth]{Figures/experiment/lime_139_fp_table.pdf}
    \includegraphics[width=0.19\linewidth]{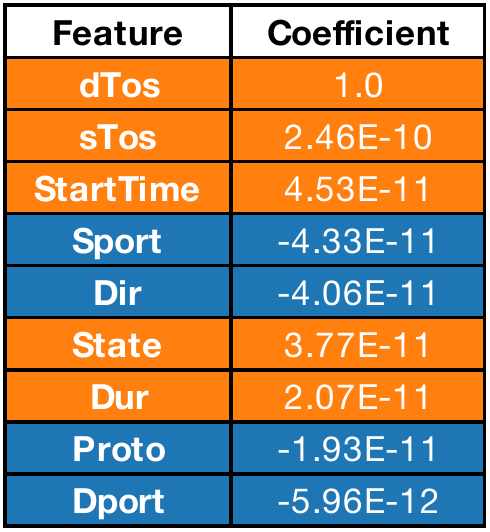}
    \caption{Post-hoc explanations for the False Positive Netflow (including spurious features). Left: SHAP, Middle: LIME, Right: LEMNA.}
    \label{fig:fp-spur}
\end{figure*}

\begin{figure*}[t]
    \centering
    \includegraphics[width=0.25\linewidth]{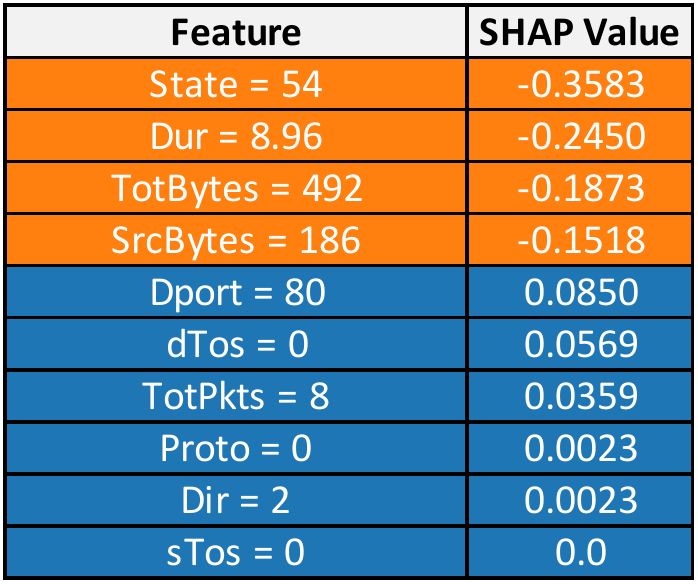}
    \includegraphics[width=0.45\linewidth]{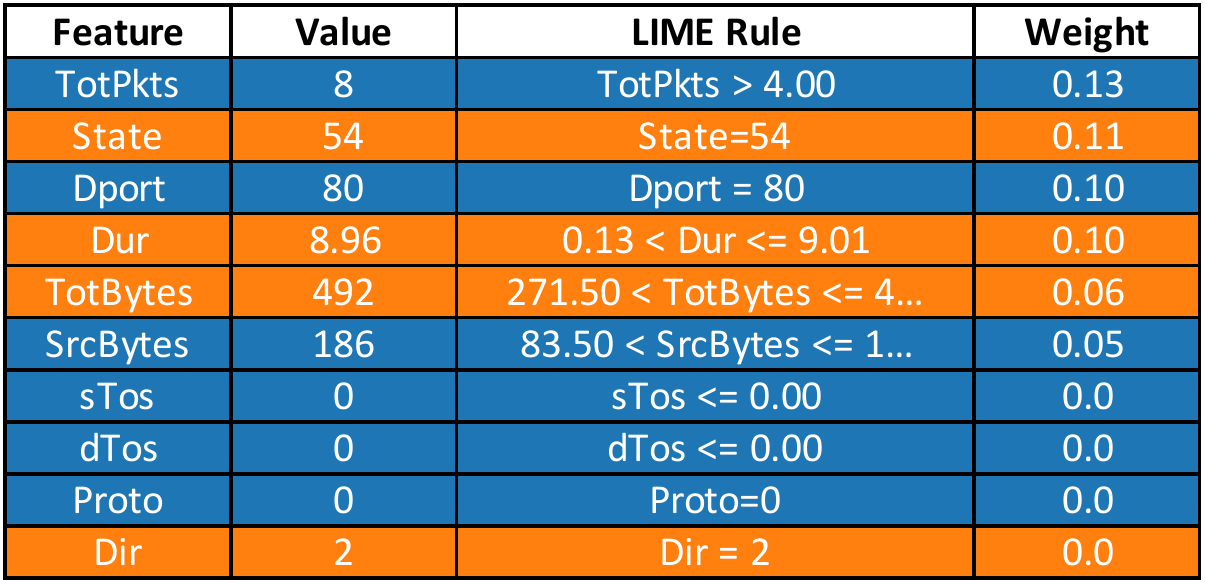}
    \caption{Post-hoc explanations for the False Positive Netflow (after removing spurious features). Left: SHAP, Right: LIME.}
    \label{fig:fp-nospur}
\end{figure*}

\begin{figure*}[t]
    \centering
    \includegraphics[width=0.23\linewidth]{Figures/experiment/shap_107_fn_table.pdf}
    \includegraphics[width=0.39\linewidth]{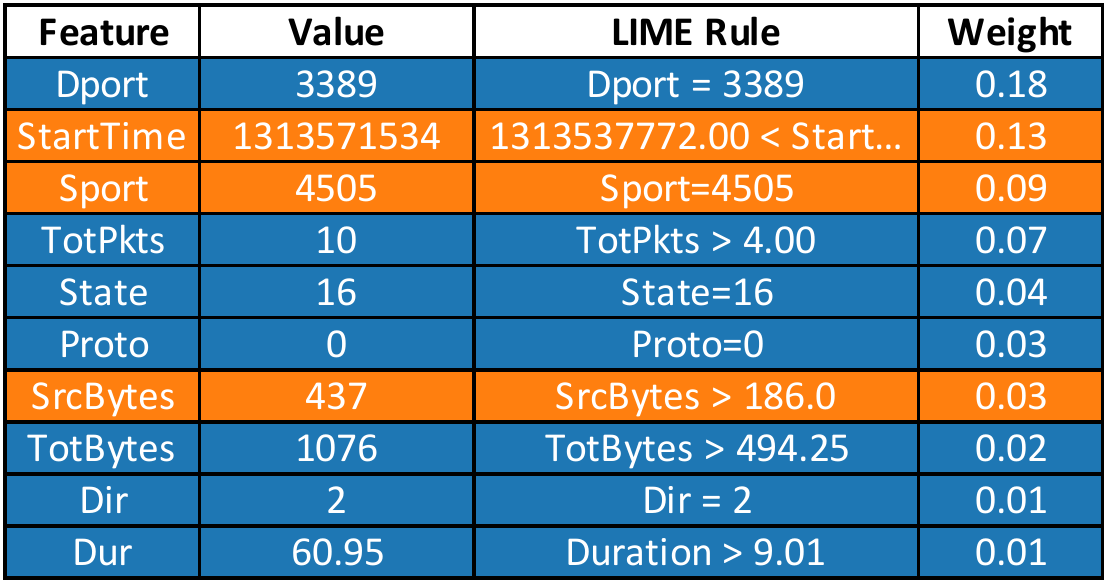}
    \includegraphics[width=0.19\linewidth]{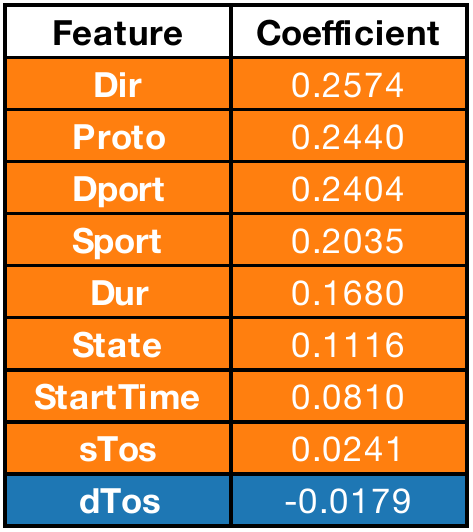}
    \caption{Post-hoc explanations for the False Negative Netflow (including spurious features). Left: SHAP, Middle: LIME, Right: LEMNA.}
    \label{fig:fn-spur}
\end{figure*}

\begin{figure*}[t]
    \centering
    \includegraphics[width=0.25\linewidth]{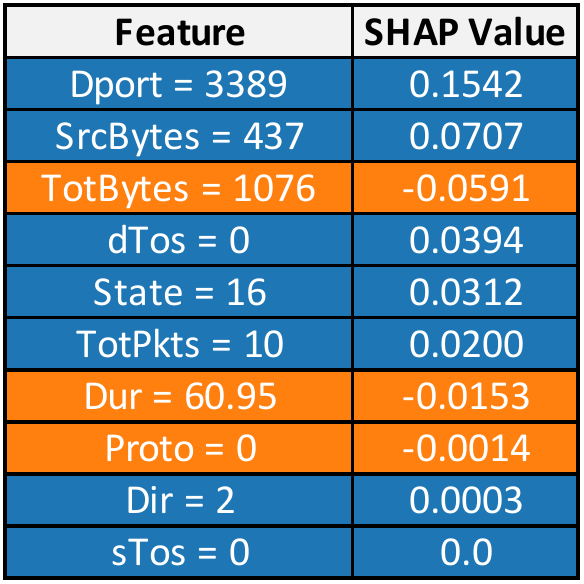}
    \includegraphics[width=0.45\linewidth]{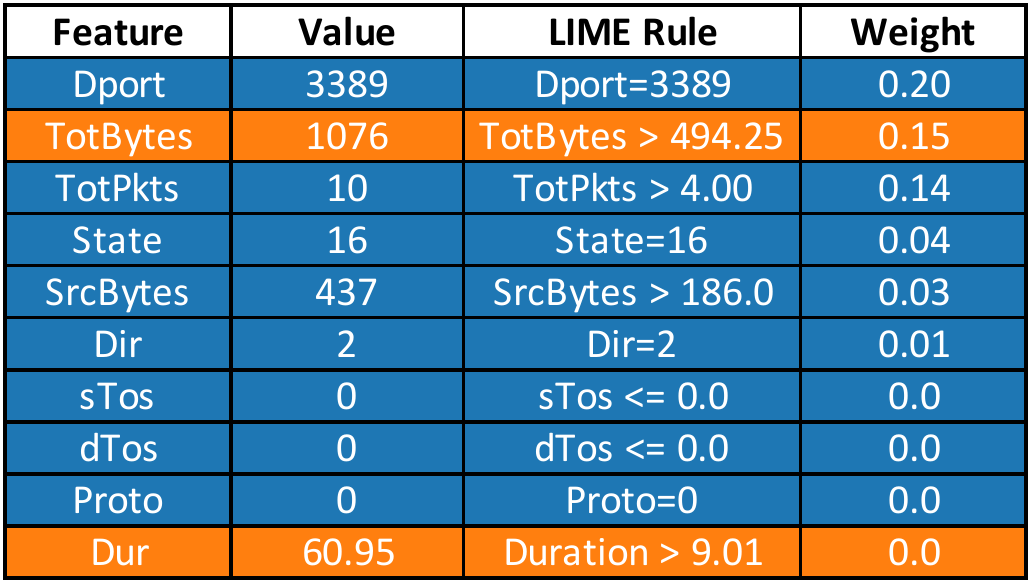}
    \caption{Post-hoc explanations for the False Negative Netflow (after removing spurious features). Left: SHAP, Right: LIME.}
    \label{fig:fn-nospur}
\end{figure*}

\section{XAI Pipeline Design}\label{pipeline-design}
We develop a modular XAI pipeline in Python (since it has in-built support for many popular models and explainers). The pipeline has three components: (1) The parser parses the input data (train and test) in either CSV or NumPy array format. The user can specify to the parser which feature fields should be read by means of providing a configuration file for the parser (2) The classifiers are implemented as a wrapper over the ML algorithms provided by \texttt{scikit-learn}. We currently support decision trees, logistic regression, explainable boosting machines, random forests, gradient boosting machines, and SVMs. The wrapper specifies the ML algorithm and its hyperparameters. (3) Similarly, the explainers are also implemented as wrapper functions and currently provide support for SHAP, LIME, LEMNA, and ELI5. The modules can be extended for added support of custom parsers, models and explainers. For the sake of reproducibility, the pipeline saves the model, predictions and explanations in a file. 

\section{LEMNA Implementation} \label{lemna-implement}
We based our implementation of LEMNA on the code by Warnecke \etal \cite{warnecke2020evaluating}. For the explanation generation, we use the following settings: $N=500, K=6, S=10$. The values of $N$ and $K$ are based on the original LEMNA paper. We do not need fused Lasso since our features do not have a temporal structure. Therefore we set $S$ to a high value, effectively turning off the fusing effect. We expect LEMNA to perform better on tabular data when using feature discretization. However, optimizing LEMNA is out of the scope of this work as we are only using existing methods for model debugging. 

\section{Explanations from the Tutorial}\label{extra-explanations}
Figure \ref{fig:global-shap} shows the global SHAP plot for the Gradient Boosting Machine (GBM) learned from the experimental dataset. 
Figures \ref{fig:fp-spur} and \ref{fig:fp-nospur} show the post-hoc explanations for the false positive Netflow with and without the identified spurious features, respectively.
Figures \ref{fig:fn-spur} and \ref{fig:fn-nospur} show the post-hoc explanations for the false negative Netflow with and without the identified spurious features, respectively.

\end{document}